\documentclass[useAMS,usegraphicx,usenatbib]{mn2e} 
\usepackage{amsmath,graphicx}
\usepackage{subfigure}
\usepackage{ulem}
\usepackage{color}
\usepackage{graphicx}
\usepackage{times} 
\usepackage{amssymb}
\usepackage{amsmath}
\usepackage{lscape}
\usepackage{url}
\usepackage{multirow}
\usepackage{longtable}

\newcommand{\km}{\rm\thinspace km}

\newcommand{\cm}{\rm\thinspace cm}

\newcommand{\cmcu}{\hbox{$\cm^3\,$}}
\newcommand{\pcmcuK}{\hbox{$\cm^{-3}\K\,$}}

\newcommand{\yr}{\rm\thinspace yr}

\newcommand{\s}{\rm\thinspace s}

\newcommand{\K}{\rm\thinspace K}

\newcommand{\Msun}{\hbox{$\rm\thinspace M_{\odot}$}}

\newcommand{\Msunpyr}{\hbox{$\Msun\yr^{-1}$}}

\newcommand{\kmps}{\hbox{$\km\s^{-1}\,$}}

\newcommand{\pcmcu}{\hbox{$\cm^{-3}\,$}}

\newcommand{\ciionefiveseven}{\hbox{[C {\sc ii}]$\lambda157\mu$m\,}}
\newcommand{\cisixzeronine}{\hbox{[C {\sc i}]$\lambda609\mu$m\,}}
\newcommand{\cithreesixnine}{\hbox{[C {\sc i}]$\lambda369\mu$m\,}}
\newcommand{\oisixthree}{\hbox{[O {\sc i}]$\lambda63\mu$m\,}}
\newcommand{\oionefourfive}{\hbox{[O {\sc i}]$\lambda145\mu$m\,}}

\def\CO{\mbox{{\rm CO}}}
\def\htwo{\mbox{{\rm H}$_2$}}

\def\h0{\mbox{{\rm H}$^0$}}
\def\he0{\mbox{{\rm He}$^0$}}

\newif\ifAMStwofonts
\AMStwofontstrue
\begin{document}
\title[Very cold clouds] {Collisional excitation of [C {\sc ii}], [O {\sc i}] and CO in Massive Galaxies}\author[R.E.A.Canning et al.] {\parbox[]{6.in}
  { R.~E.~A.~Canning$^{1,2}$\thanks{E-mail:
      rcanning@stanford.edu}, G.~J.~Ferland$^{3,4}$, A.~C.~Fabian$^{5}$, R.~M.~Johnstone$^{5}$, P.~A.~M.~van~Hoof$^{6}$, R.~L.~Porter$^{7}$, N.~Werner$^{1,2}$, R.~J.~R.~Williams$^{8}$\\ } \\
  \footnotesize
  $^{1}$Kavli Institute for Particle Astrophysics and Cosmology (KIPAC), Stanford University, 452 Lomita Mall, Stanford, CA 94305-4085, USA\\
  $^{2}$Department of Physics, Stanford University, 452 Lomita Mall, Stanford, CA 94305-4085, USA\\
  $^{3}$Department of Physics and Astronomy, University of Kentucky, Lexington, KY 40506, USA\\
  $^{4}$Centre for Theoretical Atomic, Molecular and Optical Physics, School of Mathematics and Physics, Queens University Belfast, Belfast BT7 1NN, UK\\
  $^{5}$Institute of Astronomy, Madingley Road, Cambridge, CB3 0HA\\
  $^{6}$Royal Observatory of Belgium, Ringlaan 3, 1180 Brussels, Belgium\\
$^{7}$Dept. of Physics \& Astronomy and Center for Simulational Physics, University of Georgia, Athens, GA 30602, USA\\
  $^{8}$AWE plc, Aldermaston, Reading RG7 4PR}

\maketitle
\begin{abstract}
Many massive galaxies at the centers of relaxed galaxy clusters and groups have vast reservoirs of cool ($\sim$10,000~K) and cold ($\lesssim$100~K) gas. In many low redshift brightest group and cluster galaxies this gas is lifted into the hot ISM in filamentary structures, which are long lived and are typically not forming stars. 
Two important questions are how far do these reservoirs cool and if cold gas is abundant what is the cause of the low star formation efficiency? Heating and excitation of the filaments from collisions and mixing of hot particles in the surrounding X-ray gas describes well the optical and near infra-red line ratios observed in the filaments. In this paper we examine the theoretical properties of dense, cold clouds emitting in the far infra-red and sub-millimeter through the bright lines of \ciionefiveseven, \oisixthree and CO, exposed to these energetic ionising particles. 
While some emission lines may be optically thick we find this is not sufficient to model the emission line ratios. Models where the filaments are supported by thermal pressure support alone also cannot account for the cold gas line ratios but a very modest additional pressure support, either from turbulence or magnetic fields can fit the observed [O {\sc i}]/[C {\sc ii}] line ratios by decreasing the density of the gas. This may also help stabilise the filaments against collapse leading to the low rates of star formation. Finally we make predictions for the line ratios expected from cold gas under these conditions and present diagnostic diagrams for comparison with further observations. We provide our code as an Appendix. 
\end{abstract}


\section{Introduction}
\label{intro}
The Herschel telescope opened up a wealth of lines in the far infra-red (FIR) such as the strong fine structure cooling lines of \oisixthree and \ciionefiveseven. These emission lines, which are often attributed to the presence of photo-dissociation regions (PDRs), excited by far Ultra-Violet (FUV) emission from young stars (e.g. \citealt{hollenbach1999}), have been found in a number of brightest cluster galaxies (BCGs, e.g. \citealt{edge2010a, edge2010b, mittal2011, mittal2012, werner2013}) and in X-ray and radio bright giant elliptical (gE) galaxies \citep{werner2014, guillard2014}. In these dense, X-ray bright systems the cooling time of the hot gas is short; an important consequence of which is that a heat source is required to prevent catastrophic cooling of the intracluster medium (ICM) which would lead to unprecedented growth of the galaxy. This heating is likely provided by active galactic nucleus (AGN) feedback (for a review see \citealt{fabian2012}).
\begin{figure*}
  \begin{center}
    \includegraphics[width=0.33\textwidth]{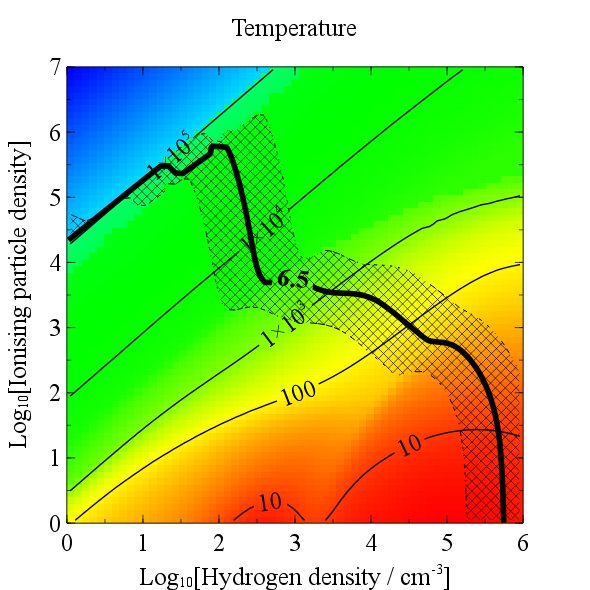}         
    \includegraphics[width=0.33\textwidth]{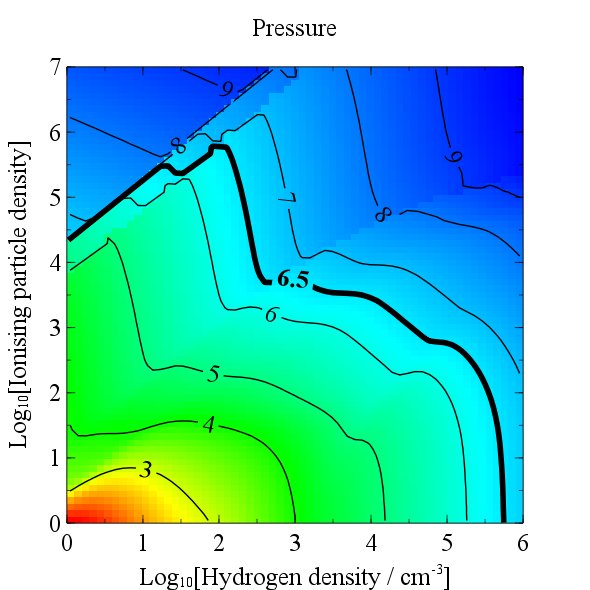}      
    \includegraphics[width=0.33\textwidth]{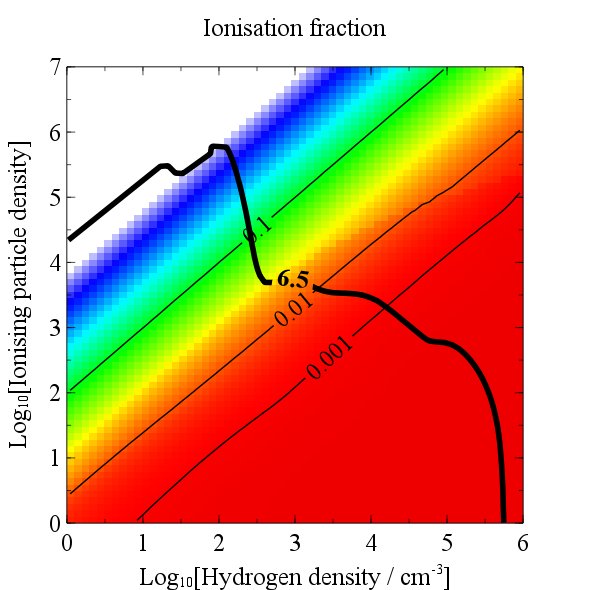}
    \caption{{\bf Left:} The temperature of the cloud. For the optically-thin case the temperature on this plot is the temperature of each `cloudlet' and in the optically-thick case it will be the surface temperature of the cloud. The solid black line represents a total pressure of 10$^{6.5}$~\pcmcuK\ (the gas pressure, derived from X-ray measurements, of the core of the Perseus cluster) and the hatched area represents pressures from 10$^{6}$-10$^{7}$~\pcmcuK. {\bf Middle:} Contours of the total gas pressure in the simple optically-thin, non-turbulent case with no additional pressure from magnetic fields. {\bf Right:} Ionisation fraction, $x_{e}$, (electron density, $n_{e}$, divided by particle density, $n$) of the gas for the same model. \label{temp_pres}}
  \end{center}
\end{figure*}

Though sample sizes are small, where extended ionised emission lines have been observed, Herschel observations of \ciionefiveseven have shown the cold (T$\lesssim$100~K) gas is also extended and coincident with filaments of ionised gas \citep{werner2013, werner2014}. \cite{werner2014} also show that the H$\alpha$/\ciionefiveseven ratio is similar implying that the ionised and cold gas share a common excitation mechanism. Though evidence for some star formation is observed in BCGs (e.g. \citealt{hubble1931, johnstone1987, mcnamara1989, allen1995, crawford1999, mcnamara2004, hicks2005, mcnamara2006, rafferty2008, odea2008, quillen2008, hicks2010, donahue2010, odea2010, canning2010a, mcdonald2012}), many of these extended filaments have no obvious star formation despite the large quantities of cool and cold gas. PDRs are therefore an unlikely heating mechanism. An important question is why are these cold gas rich systems not forming stars?

The ionised filaments are thought to be long-lived but their energy source and the reason for their quiescence has been a puzzle. \htwo, \CO, dust and polycyclic aromatic hydrocarbon (PAH) features have been shown to exist coincident with some ionised filaments (e.g. \citealt{jaffe1997, falcke1998, donahue2000, edge2001, jaffe2001, edge2002, wilman2002, salome2003,
jaffe2005, crawford2005, hatch2005, salome2006, johnstone2007, wilman2009, oonk2010, donahue2011, salome2011, canning2013}). The new information garnered from the low-ionisation and molecular emission lines seen by Herschel and the high spatial resolution made available by ALMA will be instrumental in probing density and temperature sensitive lines with which to characterise the properties of the cold gas, and in allowing the morphology and kinematics of this gas to be mapped on the same spatial scales as the optically emitting ionised filaments (e.g. \citealt{mcnamara2014, russell2014}).

\cite{ferland2008} and \cite{ferland2009} (hereafter F2009) have suggested collisions with the surrounding energetic particles as a mechanism for heating the multi-phase filaments. The authors show that their model, which allows the filaments to be described by cloudlets of varying densities, can successfully reproduce the ratios of the near infra-red (NIR) strong H$_{2}$, and optical atomic and low ionisation emission lines, which has been a struggle for photoionisation models. The model also reproduces the  characteristic signatures of relatively high He {\sc i} and [N {\sc i}] emission and of relatively low [O {\sc iii}] in the extended filaments \citep{johnstone1988, voit1997, sabra2000, hatch2005, canning2011b}, while allowing anomalously high [Ne {\sc iii}] emission, through charge-transfer reactions. However, the models are unable to match recent Herschel observations of the ratios of \oisixthree\ and \ciionefiveseven\ in the filaments. It should be noted that the authors stress their model is applicable only to optically-thin clouds and additionally that the data used in F2009 cannot constrain a large reservoir of cold clouds in the filaments as the power-law controlling the composition of the cloudlets is hinged on emission lines from gas at 10$^{4}-1000$~K.

Collisional excitations and ionisations of atomic and molecular gas by high energy particles is probably important, not only in the filaments of BCGs, but in a wide range of astrophysical environments such as the opaque molecular cores of PDRs where heating is thought to be dominated by excitations produced by cosmic rays (e.g. \citealt{hollenbach1997}), in cool and cold gas near AGNs (e.g. \citealt{shull1985}) and also in the cool gas filaments expelled in the death throes of supernovae where non-thermal electrons are generated by the hard synchrotron photoionising source (e.g. \citealt{richardson2013}).

We have extended the models of F2009 to explore the physical and chemical state of the cold (T$\lesssim$100~K) gas with densities ranging from $n=10^{0}-10^{6}$\pcmcu, and in particular we consider the effects of finite column densities of individual cloudlets. Our primary aim is to explore the discrepancy between the observed and predicted line ratios of the FIR fine structure lines of \oisixthree\ and \ciionefiveseven\ in the extended filaments of BCGs. However, our models are generally applicable to ionised, neutral and molecular gas exposed to energetic ionising particles. The spectrum from collisional heating in cool and cold gas clouds is complicated as it depends sensitively on changes in the heating and cooling of the gas which in turn depends on much of the physics of the cloud.

Section \ref{model} reviews briefly the details of the particle heating model grid and discusses the physical and chemical properties of the grid. Section \ref{additional_cold_clouds} explores the effect of a large quantity of cold clouds on the composite model spectrum for BCGs. Section \ref{optically_thick_lines} discusses the effects of optical depth on the predictions for the BCG line ratios and Sections \ref{micro_turbulence} and \ref{magnetic_fields} explore the effect of turbulence and magnetic fields. Some words of caution about the models is given in Section \ref{caution}. A discussion and predictions for future observations is presented in Section \ref{predictions}.

\section{Cold clouds in cool cores}
\label{coolcores}

As mentioned in Section \ref{intro} massive central galaxies in relaxed galaxy clusters and groups often host large amounts of cool and cold gas (e.g. \citealt{crawford1999}). The dichotomy between the lack of gas observed in central galaxies in non-cool core clusters and its abundance in cool core clusters indicates that cooling from the X-ray gas must be important in these systems (e.g. \citealt{cavagnolo2008}). However, why these extended gas reservoirs continue to shine has been an issue of contention; optical emission line ratios are typically `LINER-like' (low ionisation nuclear emission-line region). However, it has been shown that many physical processes can produce similar liner-like ratios (weak shocks e.g. \citealt{sparks1989, farage2012}; hot stars e.g. \citealt{terlevich1985}; conduction e.g. \citealt{voit1997, sparks2012}, mixing with hot gas which leads to both collisonal excitation and thermal excitation of the gas e.g. F2009; \citealt{fabian2011b} and dissipation from magnetic reconnection in magnetically supported filaments e.g. \citealt{churazov2013}).

The best, nearby, example of an extended ionised and molecular web surrounding a BCG is NGC 1275. Some regions of this extended web are forming stars but the majority are not \citep{conselice2001, canning2010a, canning2014}. Obviously, where cool gas is close to intense photoionisation from stars, such as in the inner regions of some BCGs or in star-forming filaments, this photoionisation will contribute significantly to the excitation of the gas. We must therefore look at `clean' regions, devoid of obvious star formation in order to examine the excitation mechanism for the filaments prior to their degeneration into stars, or evaporation into the hot ISM/ICM. HST SBC data shows that the SFR in a `typical' non star forming region, of NGC1275's filaments, assuming all the UV flux is due to stars, is $<$0.001~\Msunpyr \citep{johnstone2012}.

The two questions of basic importance are how much cold gas exists and if cold gas is abundant what is the cause of the low rate of star formation? Focusing on these questions we investigate the effect of energetic particles on the cold gas $\lesssim$100~K in the filaments.

\begin{figure*}
 \begin{center}
  \includegraphics[width=0.33\textwidth]{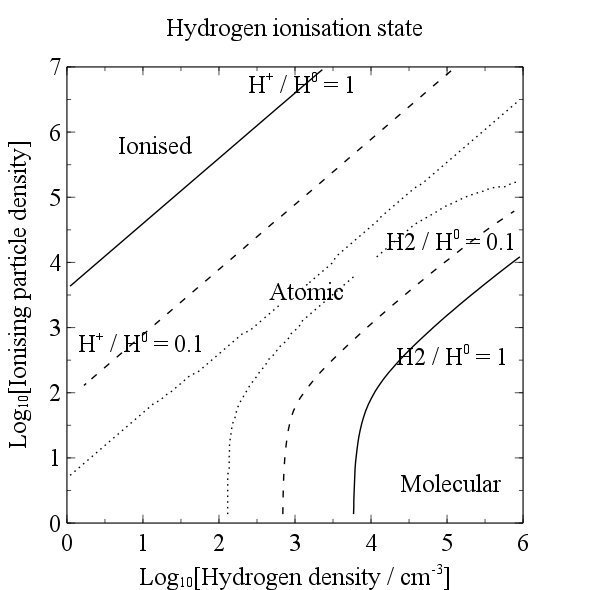}
  \includegraphics[width=0.33\textwidth]{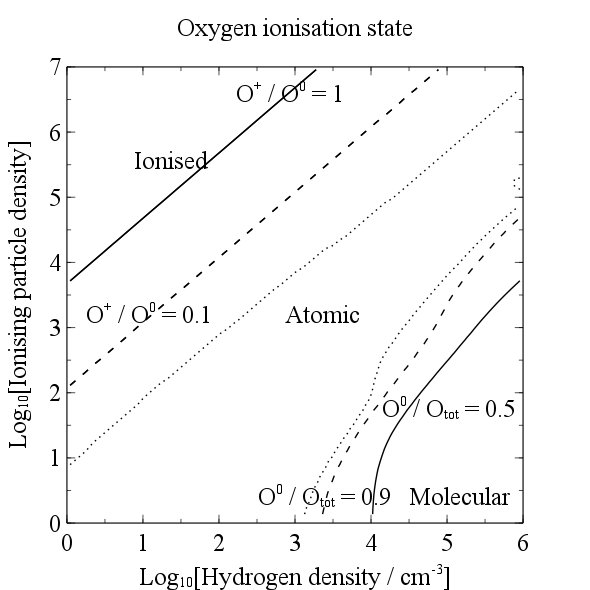}
  \includegraphics[width=0.33\textwidth]{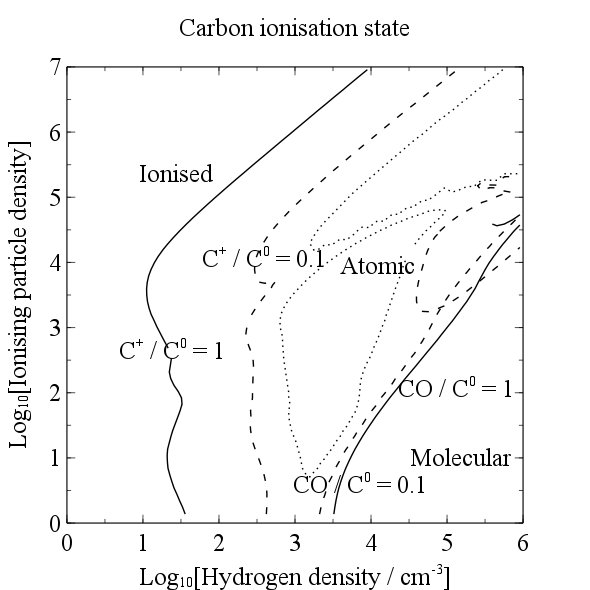}
  \caption{The ionisation structure of hydrogen, oxygen and carbon. The solid contours indicate where the atomic species and the ionised/molecular species have the same abundance. The dashed contours where the gas is 90 per cent atomic and the dotted contours where the gas is 98 per cent atomic. \label{PART_states3}}
 \end{center}
\end{figure*}

\begin{figure*}
  \begin{center}
    \includegraphics[width=0.33\textwidth]{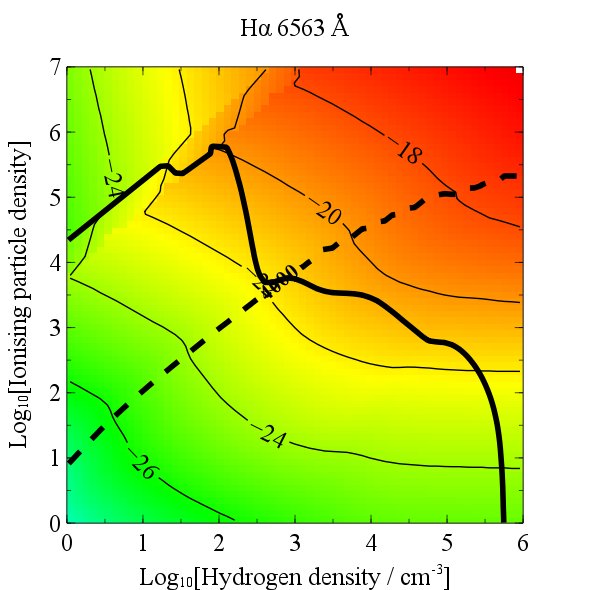}
    \includegraphics[width=0.33\textwidth]{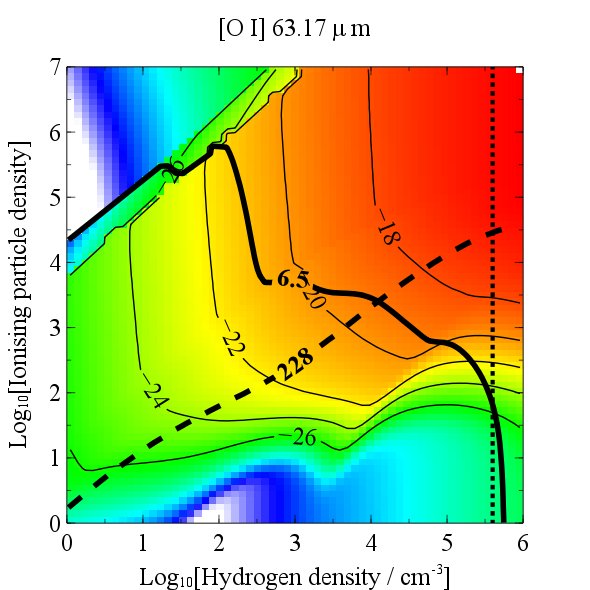}         
    \includegraphics[width=0.33\textwidth]{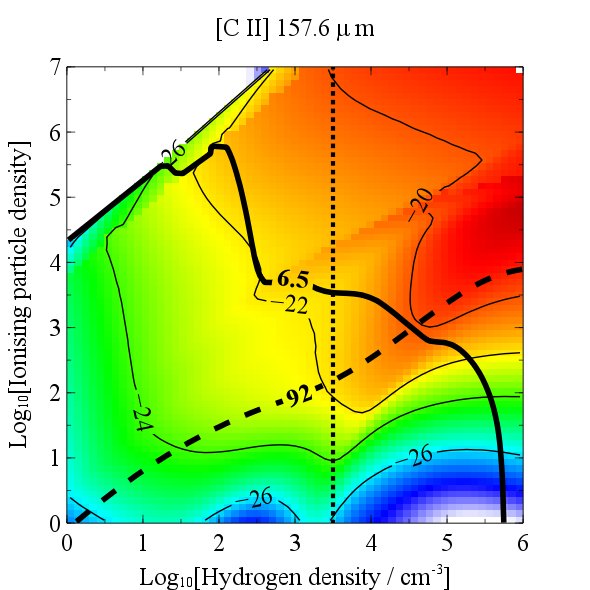}
    \caption{The log emissivity of the H$\alpha$ {\bf (left)}, \oisixthree\ {\bf (middle)}  and \ciionefiveseven\ {\bf (right)} emission lines for the simple model. The contour of constant gas pressure, relevant to the Perseus cluster (10$^{6.5}$~\pcmcuK) is drawn as a solid black line on each plot. The dashed black lines indicate the excitation temperature of the line from ground while the dotted black lines indicated the critical density of the emission line. In the optically-thin model we integrate the emissivity of each line for a given pressure over a weighted density distribution. \label{PART_OI_CII_sep}}
  \end{center}
\end{figure*}

\section{Model}
\label{model}

The particle heating model used in this paper is essentially that described in F2009 for optically-thin emission from warm ionised gas adapted to include physics pertinent to very cold clouds. We do not consider here the `extra-heating' case also considered in the aforementioned paper. The original model is motivated by spectra from extended emission line regions in BCGs which show spatially coincident emission from both molecular and ionised gas, suggesting that the total emission is from clouds with a range of densities and temperatures. F2009 parametrise the contribution of the different phases to the total volume as a weighted sum of `cloudlets' of different densities but constant pressure. The weighting function, $\alpha$, is taken to be a power-law in the cloud density. The pressure in the ionised filaments has been measured by \cite{heckman1989} and is similar to the X-ray derived pressure of the ICM.

\cite{ferland2009} use the H$\alpha$ and H$_{2}$ 12~$\mu$m emission line ratios to determine $\alpha$ for the Perseus cluster (a constant pressure of $\sim$10$^{6.5}$~\pcmcuK\ is used for the Perseus cluster). They find an $\alpha=-0.35$ can reproduce all emission lines from the warm ionised gas to within a factor of two (but see \citealt{johnstone2012} for a discussion of the H$\alpha$ flux in the Perseus cluster). This value of $\alpha$ implies the majority of the volume is filled by hot ionised gas with only a minor contribution from the more massive, cold clouds. H$\alpha$ emission is predominantly from $\sim$10$^{4}$~K gas while the H$_{2}$ 12~$\mu$m emission line is from gas $\sim$10$^{3}$~K. The authors point out that the data available at the time is not able to constrain a large reservoir of very low temperature clouds. However, extrapolation of $\alpha$ to low temperature clouds gives good agreement with mass estimates from CO emission lines though these may be optically thick \citep{salome2011}. Extrapolating this model of optically-thin lines to the very cold $\lesssim$100~K gas; the simple model predicts \oisixthree / \ciionefiveseven ratios of $\sim$20. However, the observed line ratios are $\leqq$1 \citep{mittal2011, mittal2012, werner2013, werner2014}. A myriad of physical processes could be responsible for this discrepancy in the observed ratio.

The calculations in this paper were made using C{\sc loudy} C13 (\citealt{ferland2013}) and we provide our code as Appendix A of this paper. As in F2009, due to the longevity of the filaments, we neglect regions in which the cooling is thermally unstable. We caution here that this may not be the case in the cold filaments and the line ratios in some lines are sensitive to this assumption (see Section \ref{int_limits}). \cite{chatzikos2015} studied the effects of non-equilibrium cooling but found it not to be important in the hot and warm gas.

The models shown here are of a cloud where the farthest edge from us is not open, that is photons cannot escape out the far side of the cloud. While un-physical we choose to show these models due to simplicity of explaining the underlying physics of the emission lines. We have repeated all our calculations with an open-ended cloud, the only difference in results occurs in the last zone where the temperature of the open ended cloud drops as photons can escape. This drop in temperature has the effect of slightly decreasing the emissivity in the emission lines which remain optically-thin such as H$\alpha$. No significant effect is observed in the integrated line ratios.

C{\sc loudy} contains a grain model which resolves the size distribution of the grains and can calculate the grain properties, such as the charge distribution, temperature and grain opacities separately for grains of a given radius. The grain photoelectric heating can then be determined for graphite, silicate and PAH features separately. A description of the C{\sc loudy} grain model is given in \cite{vanhoof2001, vanhoof2004}. We include dust grains with an ISM abundance and PAHs using the distribution from \cite{bakes1994} with an abundance of $10^{-4.6}$ carbon atoms per hydrogen atom in PAH features. The Jura rate is assumed for the formation of H$_{2}$ on grains and the dissociation of H$_{2}$ is calculated self consistently within C{\sc loudy}. 

In F2009 simulations are of a representative unit volume of gas. The actual filament is envisioned as being composed of many such parcels with a range of densities, temperatures, and ionizations and also with a range of column densities. For this approach to work the simulation of the unit volume is set to be optically thick in all resonance lines, as would happen if part of a much larger column density. C{\sc loudy} calculates a full model of the hydrogen atom and its emission physics.  We use this model in the predictions in this paper but assume that the Lyman lines are all optically thick.  The effect of this optical depth is to stop fluorescent pumping by the metagalactic background, and prevent Lyman lines produced in the recombination process from freely escaping.  Rather, Lyman lines undergo a large number of scatterings and are converted into Balmer and other lines, are absorbed by dust, or escape from the cloud.

The variable parameters in the particle heating model are the density of ionising particles measured relative to the Galactic cosmic ray background and the volume density of hydrogen in the gas, $n$. The energy density of the Galactic cosmic ray background is $\sim$1.8~eV~cm$^{-3}$ which corresponds to a pressure of ionising particles of $\sim$2$\times$10$^{4}$~K~cm$^{-3}$. In the centre of the Perseus cluster \cite{sanders2007} find an electron energy density $\sim$1000 times the Galactic background. We investigate the effect of ionising energetic particles over a range of values corresponding to $\sim$1 to $\sim$10$^{7}$ times the density of the Galactic cosmic ray background and gas densities of $\sim$1 to $\sim$10$^{6}$ hydrogen nuclei per cm cubed.

\subsection{Integration limits}
\label{int_limits}

\begin{figure*}
  \begin{center}
   \includegraphics[width=0.45\textwidth]{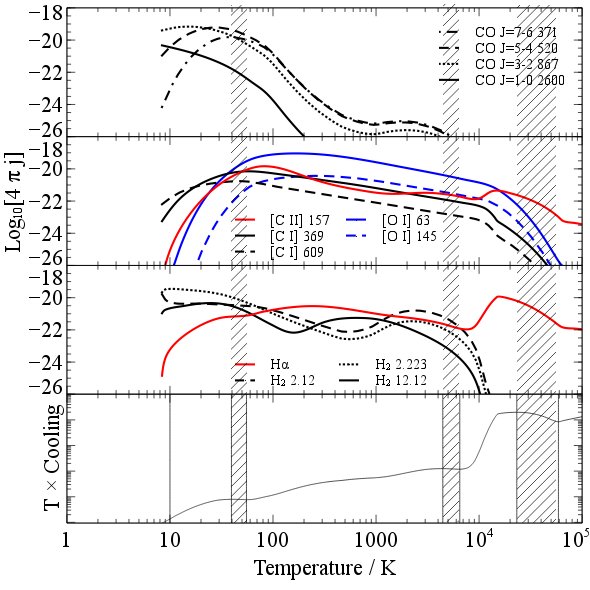}         
    \includegraphics[width=0.45\textwidth]{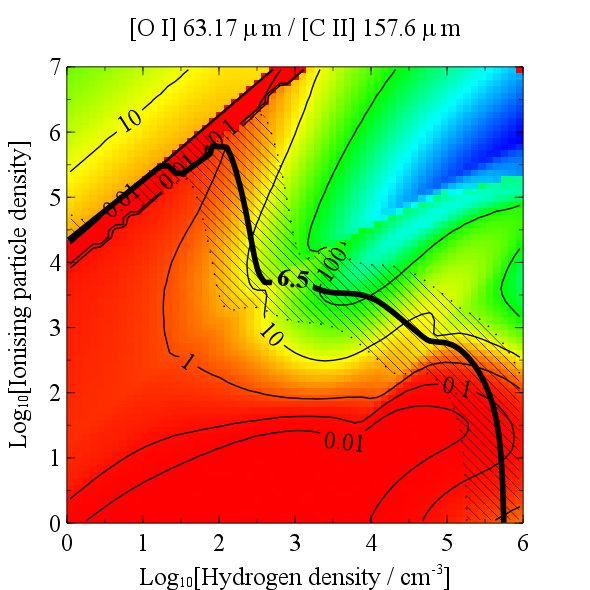}         
    \caption{{\bf Left:} The temperature dependence of the emissivity of some FIR and sub-mm cooling lines. The regions where the gas is thermally unstable are shaded. These are determined by from the cooling$\times$T curve in the bottom panel. {\bf Right:} The line ratio, \oisixthree\ / \ciionefiveseven, calculated for an optically-thin cloud for different hydrogen densities and ionising particle densities. The ratio is very sensitive to variations in the cloud properties. The contour of constant total pressure, relevant to the Perseus cluster (10$^{6.5}$~\pcmcuK) is drawn as a solid black line. The ratio increases towards the top right-hand corner of the grid. This corresponds to an increase in density at constant temperature and can be understood as the critical densities of the [O {\sc i}] lines are much larger than that of the [C {\sc ii}] lines.  \label{PART_OI_CII}}
  \end{center}
\end{figure*}

As mentioned in the previous section, the calculations in this paper neglect regions in which the cooling is thermally unstable. Thermal stability for a constant pressure gas is given by,
\begin{equation}
 \left[\frac{\partial(C-H)}{\partial~T}\right]_{P}= \left[\frac{\partial(C-H)}{\partial~T}\right]_{\rho}-\frac{\rho_{0}}{T_{0}}\left[\frac{\partial(C-H)}{\partial~\rho}\right]_{T}>0,
\end{equation}
where $C$ is the cooling rate, $H$ is the heating rate, $P$ is the pressure, $\rho$ is the density and $T$ is the temperature (\citealt{field1965}, see also Fig. 13 of F2009). The left hand panel of Fig. \ref{PART_OI_CII} shows the emissivity versus temperature profiles of H$\alpha$ and many lines from cold gas for the zero turbulence case. The bottom panel show the cooling multiplied by the temperature. The thermally unstable regions of the grid are shown by the shaded regions and the line emissivity in these regions, or equivalently cloudlets with these temperatures, are not included in the integral. 

Additionally we use a lower and upper integration limit for the temperature of the cloudlets of the lowest temperature in the grid at that pressure and 10$^{5}$~K respectively. The emission from cool and cold gas has very little emissivity beyond a few 10$^{4}$~K so the choice of upper limit of the integral has negligible effect on the total emission in the lines. The lowest temperature, at each density, occurs where there is no additional ionising particle flux. At this point the excitation in the gas is only by the background cosmic ray ionisation radiation (we assume a mean H$^{0}$ ionisation rate of 2$\times10^{-16}~s^{-1}$ from \citealt{indriolo2007}), the CMB, and the ISM photoionising radiation field. We caution that the total emission from some lines from very cold gas is sensitive to this lower limit cut-off.

\subsection{Overview of physical and chemical properties of the grid}
\label{physical_properties_of_the_grid}

Ionised and cold gas filaments in BCGs are situated in the hot ISM, surrounded by the X-ray emitting galaxy halo. The penetrating nature of X-rays means they can input energy locally into cold and dense clouds creating partially ionised regions within predominantly atomic or molecular gas. The X-ray photons action is to; one, ionise; two, excite; or three, heat the gas. The contribution of each of these mechanisms depends on both the ionisation fraction ($X_{e}=n_{e}/n$) of the gas and on its composition; whether the gas is mostly atomic or molecular. It is less sensitive to the shape of the ionising continuum, and the absolute values of the electron ($n_{e}$) or hydrogen ($n$) densities. Additionally, electrons and ions with a high kinetic temperature exist in the hot halo which can also penetrate the filaments and ionise and excite the cold gas. 

In neutral gas most of the ionisation and excitation by the supra-thermal electrons will be of H$^{0}$ due to both its larger abundance and larger cross sections for interaction compared with other neutral elements or first ions. Ionisation and excitation of He$^{0}$ and He$^{+}$ will be the next most important interactions. Primary and Auger ionisation will lead to secondary electrons which can collisionally ionise or excite other species in the gas. Excitation will generate line photons which can both interact further with the gas or may escape the cloud causing the gas to cool. Finally, direct heating of the gas is achieved through Coulomb collisions (see \citealt{maloney1996} for an in-depth description of X-ray irradiation of cold dense clouds). Here we are modelling the effect of collisional excitation with energetic particles which may arise from energetic particles in the hot halo interacting with the cold cloud or energetic electrons from X-ray irradiation of the cloud. 

The temperature, total pressure and ionisation fraction of the gas are shown in Fig. \ref{temp_pres} for the simple, optically-thin particle heating model; no additional pressure from turbulent motions or magnetic fields are included. In all three plots the thick black contour indicates a gas pressure of $\sim$10$^{6.5}$~\pcmcuK, appropriate to the Perseus cluster of galaxies. As expected, as we increase the flux of ionising particles, for fixed density, the temperature and pressure increase and as does the ionisation fraction. For a fixed flux of ionising particles, increasing the density of the cloudlet corresponds to a decrease in the temperature and an increase in the pressure. As the temperature decreases so does the ionisation fraction. 

Where the ionisation fraction ($X_{e}=n_{e}/n$) is greater than a few percent, Coulomb collisions with the thermal electron population dominate the interactions of the supra-thermal electrons and provide the majority of the heating in the cloud. When $X_{e}$ falls below 1 per cent ionisations of neutral hydrogen dominate with excitations also becoming increasingly important \citep{shull1985, maloney1996}. The ionisation structure of hydrogen, oxygen and carbon, for a range of ionising particle densities and hydrogen densities is shown in Fig. \ref{PART_states3}. H$^{0}$ persists over much of the parameter space with the gas becoming fully ionised at several tens of thousands of Kelvin and fully molecular at only a few tens. In atomic gas, when the energy of the electrons drops below the smallest excitation energy, all of the remaining energy must go into heating the gas. Therefore, for a gas dominated by neutral hydrogen, once the supra-thermal electron energy falls below 10.2 eV, Coulomb heating is the only heating process. So a low ionisation fraction in atomic gas leads to a low heating efficiency \citep{shull1985, xu1991}.

At high densities and low temperatures the gas transitions to a primarily molecular phase, though pockets of ionised and atomic regions will remain. Heating is more efficient in molecular gas and is dominated by collisional de-excitation of vibrationally excited H$_{2}$ molecules and by H$_{2}$ dissociation by photons and ionisation by secondary electrons. Significant amounts of energy can also go into the excitation of the rotation-vibration bands of H$_{2}$ and into the dissociative electronic states \citep{glassgold1973, voit1991}.  In Fig. \ref{temp_pres} the temperature increases around a density of 10$^{3.3}$~\pcmcu. This is due to the increase in the H$_{2}$ gas fraction leading to a higher heating efficiency.

\begin{figure*}
  \begin{center}
    \includegraphics[width=0.4\textwidth]{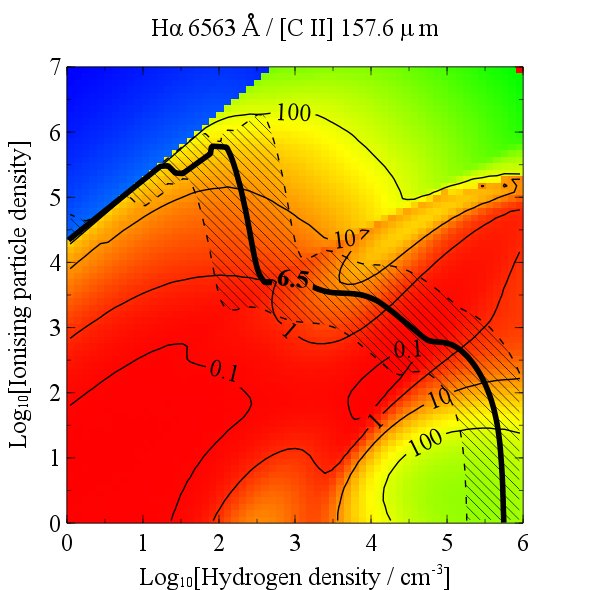}         
    \includegraphics[width=0.4\textwidth]{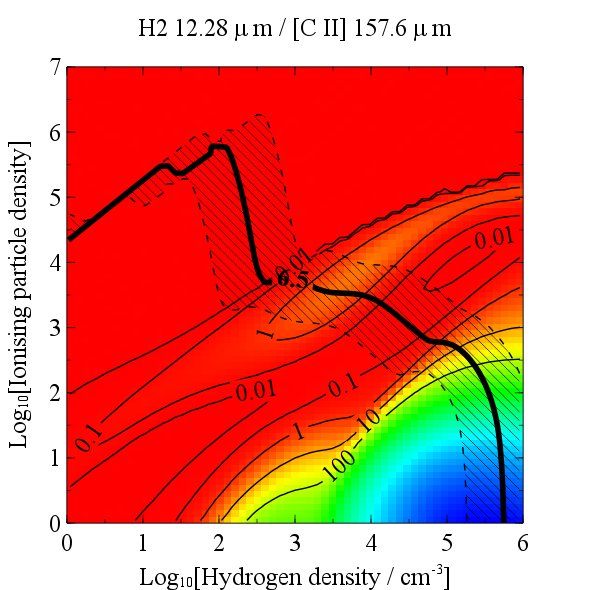}        
    \caption{The ratio of the emissivity of H$\alpha$ to that of \ciionefiveseven\ ({\bf left}) and the ratio of the emissivity of H$_{2}$ 12.28 $\mu$m to that of \ciionefiveseven\ ({\bf right}) for optically thin clouds with the same conditions as in Fig. \ref{PART_OI_CII}. Observationally H$\alpha$/\ciionefiveseven\ $\sim$1.3 \protect \citep{mittal2012, werner2014} and H$_{2}$ 12$\mu$ m/\ciionefiveseven $\sim0.025$ (F2009). \label{PART_CII_HaCII_H2CII}}
  \end{center}
\end{figure*}

Fig. \ref{PART_OI_CII_sep} shows the variation in the emissivity of the H$\alpha$, \oisixthree and \ciionefiveseven emission lines respectively across the whole parameter space. The dashed line in each case indicates the excitation temperature of the line from the ground while the dotted black lines indicated the critical densities (for collisions with electrons) of the lines; $n_{cr}^{[\mathrm{O~I}]}=5\times10^{5}$~\pcmcu and $n_{cr}^{[\mathrm{C~II}]}=3\times10^{3}$~\pcmcu. However, it should be noted electrons are not the only colliders in the gas.

Collisional excitation is the dominant contributor to the line emission in all regions of the plot. However, above the excitation temperature excitation is by both thermal particles and secondary electrons whilst below the emissivity is dominated by excitation with secondary electrons. In our model H$\alpha$ emission in the filaments arises from collisions with both populations (see upper panel of Fig. 15 of \citealt{ferland2009}), and the geometry of the resulting clouds will be permeated by pockets of high and low temperature in a `swiss cheese-like' fashion. The degree of polarisation expected from the model is therefore complex (see also \citealt{sparks2014}). 

The emissivity of the collisionally excited lines is essentially proportional to the $n_{a}n_{b}f(T)$ where $n_{a}$ and $n_{b}$ are the species colliding and $f(T)$ is the Boltzmann factor of the line for the case where collisions are occurring with thermal particles and is a constant for the case where the collisions are occurring with supra-thermal secondary electrons. Therefore, in the lower right hand corner of the plots, where the gas is mostly molecular and the temperature is too low for collisional excitation by thermal particles, at constant density the emissivity will increase by a more-or-less constant amount which is proportional to the number of secondary electrons available to collide with. The approximate flattening of the emissivity for constant ionising particle density with increasing hydrogen density is due to the high abundance of targets, such as H$^{0}$, compared with the low abundance of ionising secondaries; increasing the density doesn't affect the emissivity. The supra-thermal particles prevent the gas from becoming fully molecular even in the densest regions. 

As the temperature of the gas increases and we move towards the upper left hand side of the plot collisions by thermal particles also become important and the line emissivity is less sensitive to the number of ionising particles. Significantly above the excitation temperature the emissivity remains essentially constant with increased gas temperature until the temperature is high enough that most atoms are ionised and the line emissivity drops.

\section{Complicating the model}
\label{complicated}

\subsection{Additional cold clouds}
\label{additional_cold_clouds}

The formalism presented in F2009 describes the cloud in terms of a weighted sum of very small `cloudlets' of different density phases but at constant pressure. The weighting function is parametrised as a fraction of the total volume filled by each phase and an assumption is made that the function takes the form of a power-law in density. The total emissivity is found by integrating the density dependent emissivity over the cloud distribution. 

In reality we do not know the shape of the cloud distribution which may be highly complex; in practice a power law is chosen for simplicity. Physically, the variation of the power-law index ($\alpha$) tells us the contribution to the total emissivity in a particular emission line, from gas of different temperatures, thus a relatively high $\alpha$ would show regions where dense, cold gas dominates over warmer less dense regions. As the cloud is assumed to maintain a constant gas pressure a simple relation exists for the temperature and total particle density.

In the right hand panel of Fig. \ref{PART_OI_CII} the ratio of \oisixthree\ over \ciionefiveseven\ is predicted for a grid of ionising particle fluxes and cloud densities. Our assumption of constant pressure constrains us to lie along the black contour in the right hand panel of Fig. \ref{PART_OI_CII}. The hashed region shows where contours of constant pressure of 10$^{6}$~\pcmcuK\ to 10$^{7}$~\pcmcuK\ lie. If a power law is not an appropriate parametrisation for the coldest clouds then their predicted ratios will be different. For example, at a pressure of 10$^{6.5}$~\pcmcuK\, in gas with a temperature of T$<$25~K ($n>10^{5.1}$) \ciionefiveseven\ dominates over \oisixthree\ emission while at higher temperatures the contrary is true.

Making the assumption that the power law is not an appropriate parametrisation for the coldest gaseous phases the most efficient way to produce a low ratio of \oisixthree\ over \ciionefiveseven\ would be by assuming additional cold clouds where the ratio is very low. The left hand panel of Fig. \ref{PART_OI_CII} shows for a constant pressure cloud with 10$^{6.5}$~\pcmcuK, the \oisixthree\ over \ciionefiveseven\ ratio is lowest ($\sim0.05$) for hydrogen densities of $\sim$10$^{5.6}$\pcmcu\ and where the ionising particle density is of order 10$^{1.5}$ times the Galactic cosmic ray background.

Assuming a cloud distribution with the power law function described in \cite{ferland2009} and an additional delta function where the ratio of \oisixthree\ over \ciionefiveseven\ is lowest we require the emissivity of this additional cold cloud to be a factor of $1.8\times10^{4}$ times that expected for a 1\cmcu\ cloud with the same properties, in order for the \oisixthree\ over \ciionefiveseven\ ratio to be $\sim$1. However, the addition of more cold clouds has a significant effect on the other line ratios. Importantly H$\alpha$ and H$_{2}$ 12.28 $\mu$m are significantly stronger for high density, low temperature clouds than \ciionefiveseven\ by factors of $\sim$100 and $\sim$20,000 times respectively, which is not observed in the filaments (see Fig. \ref{PART_CII_HaCII_H2CII}).

\subsection{Optically thick lines}
\label{optically_thick_lines}

The majority of the lines of interest in F2009 are optically-thin thus validating the use of a unit volume of gas in the predictions of the line ratios. However, as previously mentioned, some lines in the wavelength range of Herschel and ALMA may become optically-thick at reasonable column densities and as such require special attention.

The intensity of radiation from a cloud decreases along its path due to scattering and absorption and increases due to spontaneous and stimulated emission. For an optically-thin plasma, all radiation generated within the plasma is able to leave it, spontaneous emission is all that is required and the matter is not in thermal equilibrium with the radiation. Hence an optically-thin source will radiate below the black body limit. In an optically-thick system photons are absorbed and re-emitted many times before leaving the cloud and the cloud is considered to be at a quasi-equilibrium temperature where emission from the cloud in all directions will be the same. This sets an upper limit to the intensity obtained at a specific frequency from a thermal source at a specific temperature and is the black body limit. The luminosity in the line, per steradian, at the black body limit can be estimated by,
\begin{equation}
L_{line}=T_{b}f_{v}\Delta V\pi R^{2}=\left(\frac{h\nu}{k_{B}}\right)\frac{1}{e^{\frac{h\nu}{k_{B}T}}-1}f_{v}\Delta V\pi R^{2}
\label{equone}
\end{equation}
where $\Delta V$ is the velocity width of the line in \kmps, due to micro turbulence, $f_{v}$ is the velocity filling factor, and $T_{b}$ is the line's brightness temperature in K. For \oisixthree\ and \ciionefiveseven\ we cannot make the simplifying assumption that $h\nu\ll kT$.

\cite{mittal2011} have observed [C {\sc ii}]$\lambda$157$\mu$m and [O {\sc i}]$\lambda$63$\mu$m line emission using Herschel and measure line widths of a few hundred \kmps. The flux expected at the black body limit is therefore $\sim$1$\times$10$^{-16}$ W~m$^{-2}$ for [C {\sc ii}] and $\sim$6$\times$10$^{-15}$ W~m$^{-2}$ for [O {\sc i}] (with a filling factor of 1). The measured flux of \ciionefiveseven\ is comparable to that of the black body limit while the measured flux of \oisixthree\ is a factor of one hundred lower.

This is a very conservative limit as the real velocity width of the line is likely to be a lot smaller. \cite{salome2008a} measure line widths of the CO $J=2-1$ line in NGC 1275 in the Perseus cluster with the 2.5'' beam of the Plateau de Bure Interferometer, to be as low as $\sim$30 \kmps. The smallest Herschel angular resolution is 5'', while the cool ionised gas filaments are determined, with the Hubble Space Telescope, to have an upper limit of only 0.2 arcsec (70 pc) wide in Perseus. It is therefore very likely, within this large aperture, that the velocity width is dominated by the macro turbulence of many intertwined filaments, not the micro turbulence of the emission line.

\begin{figure}
  \begin{center}
   \includegraphics[width=0.45\textwidth]{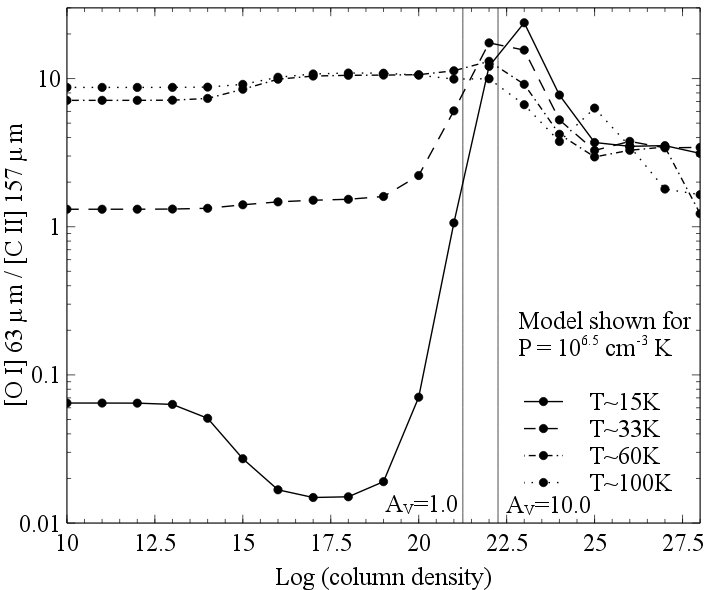} 
   \includegraphics[width=0.45\textwidth]{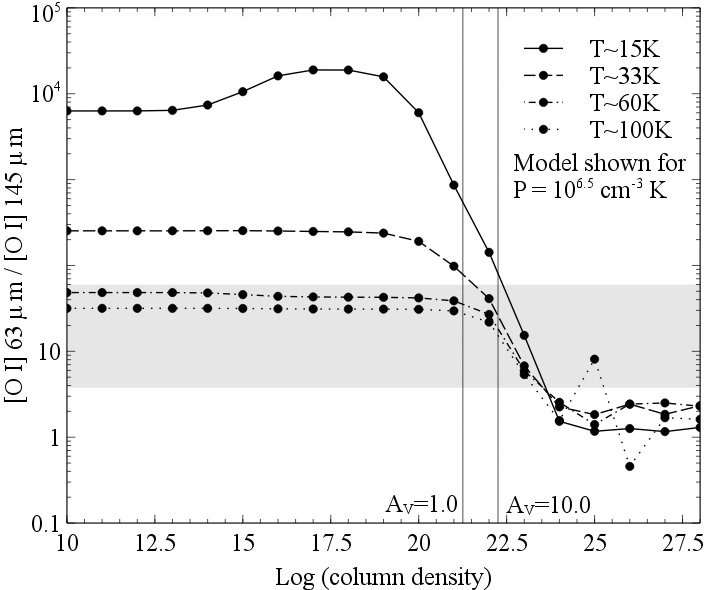} 
   \caption{The predicted \oisixthree\ / \ciionefiveseven\ ({\bf top}) and \oisixthree / \oionefourfive\ ({\bf bottom}) line ratios versus column density for cloudlets at a constant pressure of $\sim10^{6.5}$~\pcmcuK, but with varying surface temperatures. At high optical depths the \oisixthree\ / \ciionefiveseven\ ratio tends to a few, this is larger than is observed. Observational \oisixthree\ / \ciionefiveseven\ ratios in the filaments are typically less $<$1. In the bottom plot the grey region shows the currently detected values for the \oisixthree / \oionefourfive\ line ratio, we note that in many cases the \oionefourfive\ emission line is too weak to detect. These line ratios favour optically thin gas at `warm' temperatures (50-100~K).   \label{oi_cii_depth}}
  \end{center}
\end{figure}

To explore the effect of high column on the predictions of the particle heating model we will first make the simplifying assumption that the cloud is at one initial temperature and model the emergent line intensities of important IR cooling lines for clouds of different column density. In this manner we are studying conditions across a finite column density cloud with surface properties corresponding to a single point in Fig. \ref{temp_pres}. We show in Fig. \ref{oi_cii_depth} the emission line ratios of \oisixthree/\ciionefiveseven\ and \oisixthree/\oionefourfive\ for four such models with surface temperatures of 15~K, 33~K, 60~K, and 100~K at a pressure of $\sim10^{6.5}$~\pcmcuK, the emissivity profiles of many cold gas lines are shown in Fig. \ref{emissivity_column_const_press}.

Fig. \ref{oi_cii_depth} shows that for a constant pressure cloud, even at low initial temperature, \oisixthree/\ciionefiveseven\ $>1$ once the lines are optically-thick. This is expected from the black body limit, assuming the emission lines come from the same gas, given in Equ. \ref{equone}. If both lines are saturated we expect \oisixthree/\ciionefiveseven\ $\sim$3, therefore under equilibrium conditions and assuming no additional pressure support in the gas, an optically-thick gas is not sufficient to explain the observed line ratios of \oisixthree\ and \ciionefiveseven\ in the filaments.

The \oisixthree and \oionefourfive emission lines should be produced from the same clouds so this ratio has less degeneracies than the \oisixthree/\ciionefiveseven\ ratio and can be a more sensitive test of the optical depth in the lines. In the optically thick limit, for a reasonable range of temperatures, Fig. \ref{oi_cii_depth} shows the ratio of \oisixthree/\ciionefiveseven\ should be lower than has currently been observed (observations are indicated by the grey region). We note that in many cases \oionefourfive\ is too weak to be detected and as such the upper limit to the grey region is likely to increase with deeper observations. In addition, in the no-turbulence case for lines with temperature $\lesssim$100~K the [O I]63$\mu$m line becomes optically-thick around a column density of $\sim10^{22}$~cm$^{-2}$, corresponding to a high extinction of A$_{\mathrm{V}}=$5.6 (assuming N$_{\mathrm{H}}$=1.8$\times$10$^{21}$A$_{\mathrm{V}}$ from \citealt{predehl1995}) The \ciionefiveseven and \oionefourfive lines become optically thick at columns which are prohibitively large (see also \citealt{liseau2006}).

\begin{figure*}
  \begin{center}
   \includegraphics[width=0.33\textwidth]{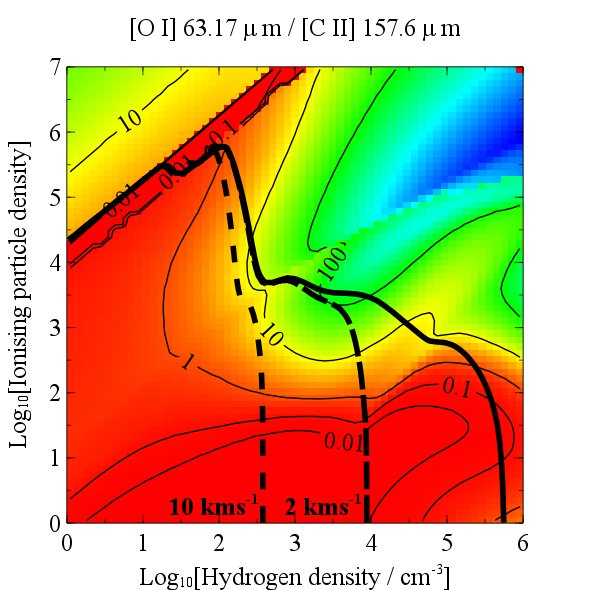}    
   \includegraphics[width=0.33\textwidth]{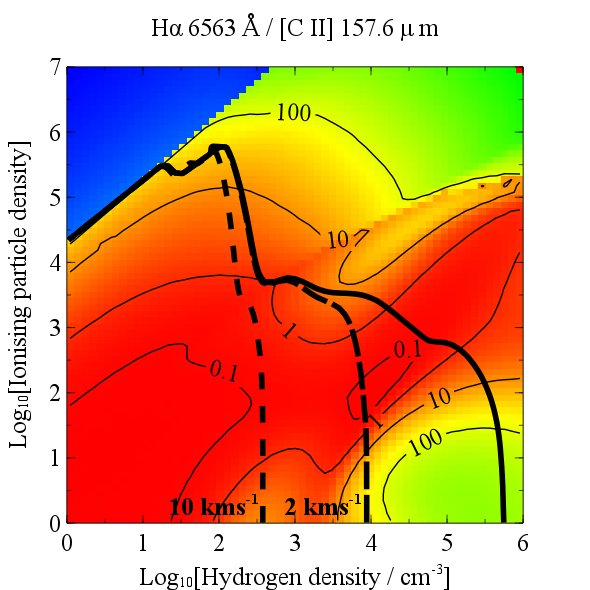}  
   \includegraphics[width=0.33\textwidth]{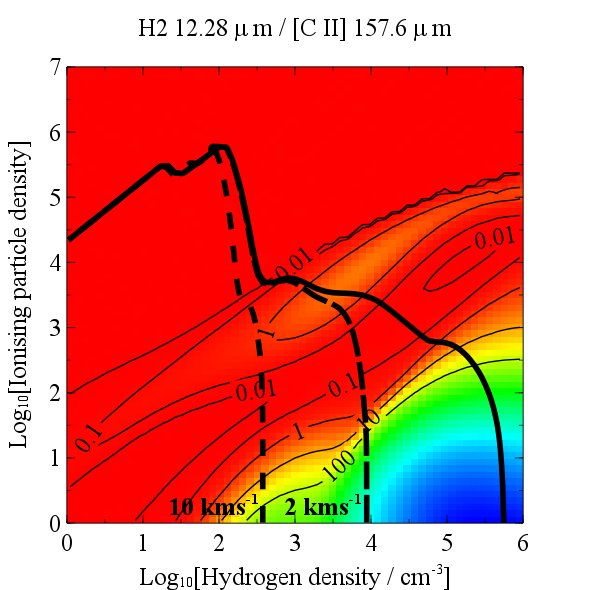}    
  \caption{The ratios of key lines overlaid with contours of gas pressure in the cases where none, 2~\kmps\ and 10~\kmps\ turbulence is present in the gas. {\bf Left:}  \oisixthree / \ciionefiveseven. {\bf Middle:}  H$\alpha$ / \ciionefiveseven. {\bf Right:} H$_{2}$ 12.28 $\mu$m / \ciionefiveseven. The measured fluxes are the integral of the cloudlets, indicated by the pressure contour. Observationally the fluxes are \oisixthree / \ciionefiveseven $\lesssim1$, H$\alpha$/\ciionefiveseven\ $\sim$1.3 \protect \citep{mittal2012, werner2014} and H$_{2}$ 12$\mu$ m/\ciionefiveseven $\sim0.025$ (F2009). \label{turb_figs}}
  \end{center}
\end{figure*}

\subsection{Turbulence}
\label{micro_turbulence}

In the previous sections, and in the F2009 model, no turbulence was included in the lines. We currently only have upper limits on the turbulent motions of a single cloud in the extended filaments of BCGs as the beam size is large compared with the filament threads. Optical and NIR observations of the filaments typically measure line widths, of gas at $\sim$1,000 and 10,000~K, to be between $50-200$\kmps \citep{lim2012} while \cite{salome2008a} have measured line widths of $\sim$30 \kmps, in the cold gas, using the 2.5'' beam of the Plateau de Bure Interferometer in the extended filaments of NGC 1275. The 2.5 arcsec beam is $\sim$12 times the upper limits on filament thread width determined by the Hubble space telescope \citep{fabian2008}. All these observations currently suffer from confusion of filament threads.

In Galactic giant molecular clouds line widths are typically measured as a couple of \kmps\ in the molecular gas with a measurements of a few \kmps\ in the cold and warm neutral mediums \citep{hennebelle2013}. Under 100~K the sound speed is already less than 2~\kmps\ while in $<30$~K gas the sound speed is below 1~\kmps\ so molecular clouds may be turbulent in the molecular and possibly cold neutral phases. \cite{kritsuk2011} show that scaling relations found between the $^{12}$CO(J=1-0) line-width and size and mass and size of molecular clouds \citep{larson1981} can be interpreted as a signature of supersonic motion. Other supporting evidence for supersonic motions come from the measurement of a log normal distribution of the column density \citep{vazquezsemadeni1994}. However, while turbulence predicts a log normal distribution it is not necessary that a log normal distribution implies turbulence \citep{tassis2010}.

Micro-turbulence increases the line width which suppresses the optical depths and alters the importance of shielding and pumping of lines. Self-shielding becomes less important while the ability to absorb a larger part of the continuum increases the importance of florescence. Our initial grids, of the emissivities of the emission lines, shown in Fig. \ref{PART_OI_CII_sep} are not affected by the addition of turbulence as in this simple model all lines are assumed optically-thin. We note here our cloudy models do not include shock heating of the gas.

As well as altering the chemical balance, the introduction of microturbulence affects the pressure balance of the cloud. C{\sc loudy} includes the turbulence as a velocity in \kmps, $u_{turb}$. The additional energy density is therefore $P_{turb}=\frac{F}{6}\rho u^{2}_{turb}$, where $\rho$ is the gas density and $F$ is a constant which accounts for how ordered the turbulent velocity field is. We assume $F=3$, appropriate for isotropic turbulent motions but note that the filaments are likely threaded with magnetic fields which would influence the turbulent motions of ions \citep{heiles2005}.

The effect of the additional pressure term significantly alters the predicted line ratios, from cold gas, for reasonable values of $u_{turb}$. The gas pressure contour required to keep the filament in pressure equilibrium with the surrounding hot gas is reduced as shown in Fig. \ref{turb_figs}. Essentially, when the turbulent pressure is large the gas cools at constant density. The turbulent pressure is density dependent and as such the largest deviation from the no-turbulence curve, for a constant turbulent velocity at all temperatures, is in the densest gas phases. 
As can be seen in Fig. \ref{turb_figs} this will alter the ratios of \oisixthree / \ciionefiveseven, pushing them to lower ratios while leaving those of higher temperature lines unchanged.

Turbulent dissipation would also lead to heating of the gas. We can estimate the contribution of the turbulent heating to the gas line luminosity using,
\begin{equation}
L_{turb}=\frac{3 M_{tot} v_{turb}^{3}}{2 l}~ergs~s^{-1},
\end{equation}
where $M_{tot}$ is the gas mass, $v_{turb}$ is the turbulent velocity and $l$ is the injection scale of the turbulence. We assume the very conservative values of $M_{tot}\sim10^{6}M_{\odot}$ and $v_{turb}\sim10$\kmps. The scale is the largest unknown, not least because individual filament threads have not been resolved, and so we assume here $l\sim 1/4$ HST filament width limit $\sim$ 17~pc. These values lead to $L_{turb}~\sim 5.5 \times10^{37}$~ergs s$^{-1}$. In the horseshoe filament chosen as a typically example filament by F2009 in NGC 1275, the authors find a H$\alpha$ luminosity of $7\times10^{39}$~ergs~s$^{-1}$. The \ciionefiveseven / H$\alpha$ ratio is $\sim$0.8 so $L_{[C II]}\sim5.6\times10^{39}$~ergs~s$^{-1}$. The turbulent heating is therefore only one percent of the line luminosity. A turbulent velocity of $\sim$45~\kmps\ is required to reach the \ciionefiveseven\ line luminosity, or alternatively $l~\sim~0.17~pc$. However, this still neglects the luminosity of H$_{2}$, CO and other important lines.

\subsection{Magnetic fields}
\label{magnetic_fields}

The presence of strong magnetic fields in the filaments of BCGs has been inferred from optical observations of the geometry and widths of single threads of extended filaments \citep{fabian2008}, from radio observations \citep{taylor2006} and implied from density arguments \citep{werner2013}. However, magnetic fields have never been directly detected in the filaments.
If these strong fields exist they will increase the energy density, $P_{mag}=\frac{B^{2}}{8\pi}$, also decreasing the contribution to the total pressure budget from gas pressure. 

\begin{figure*}
  \begin{center}
   \includegraphics[width=0.33\textwidth]{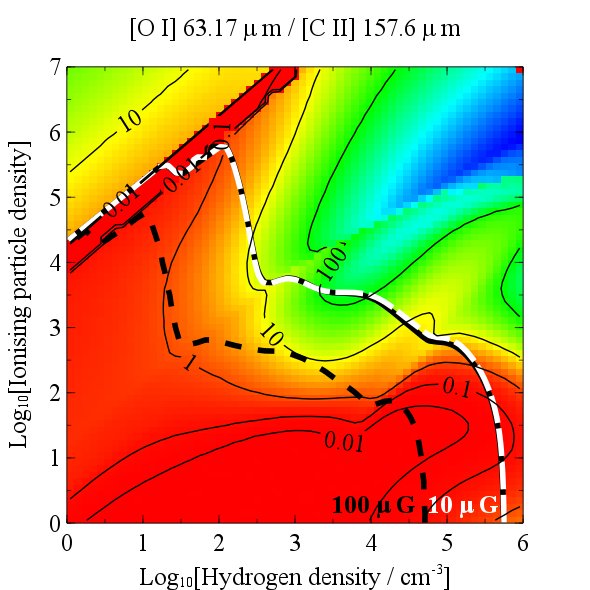}    
   \includegraphics[width=0.33\textwidth]{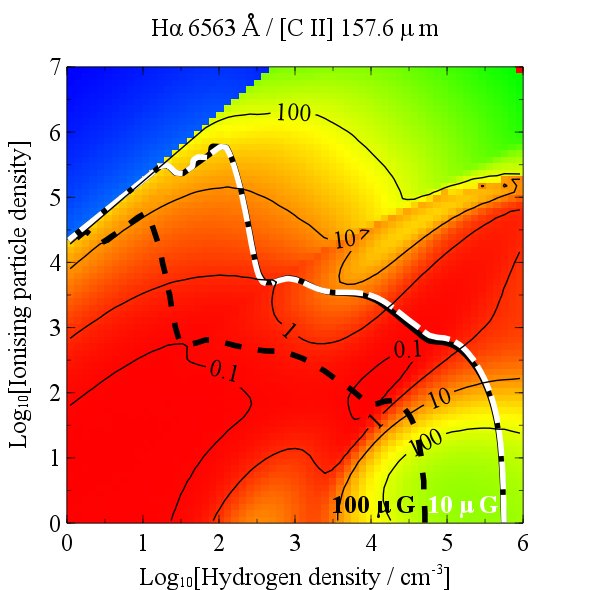}     
   \includegraphics[width=0.33\textwidth]{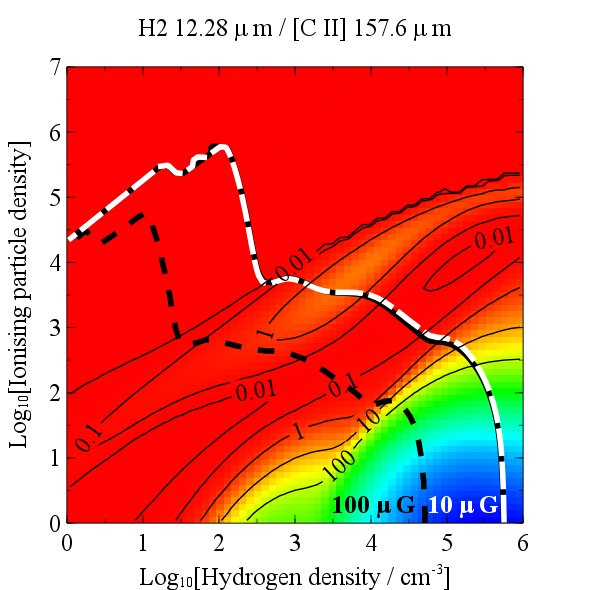}    
  \caption{The ratios of key lines overlaid with contours of gas pressure in the cases where none, 10~$\mu$G and 100~$\mu$G fields are present in the gas. {\bf Left:}  \oisixthree / \ciionefiveseven. {\bf Middle:}  H$\alpha$ / \ciionefiveseven. {\bf Right:} H$_{2}$ 12.28 $\mu$m / \ciionefiveseven. The measured fluxes are the integral of the cloudlets, indicated by the pressure contour. Observationally the fluxes are \oisixthree / \ciionefiveseven $\lesssim1$, H$\alpha$/\ciionefiveseven\ $\sim$1.3 \protect \citep{mittal2012, werner2014} and H$_{2}$ 12$\mu$ m/\ciionefiveseven $\sim0.025$ (F2009). \label{B_figs}}
  \end{center}
\end{figure*}

In our implementation $\underline{\mathrm{B}}$ is a constant and not dependent on density. This may not be the case but a full treatment is beyond the scope of the paper. As the $\underline{\mathrm{B}}$ field is not density dependent the additional pressure is a constant in all cloudlets and as such the required gas pressure is less at all densities. This is shown graphically by the contours on Fig. \ref{B_figs}. Importantly, as in the case of additional turbulent support, the required density of the coldest gas is less dense than in the fiducial model which decreases the predicted ratios of \oisixthree\ over \ciionefiveseven. However, this implementation will effect all line ratios; the optical and IR lines from the 10,000~K$-$1,000~K gas as well as the FIR and sub-mm lines from much colder gas.

\subsection{Words of caution}
\label{caution}

The models presented above explore the effect of the relaxation of certain assumptions, namely, the assumption of a power-law in density, optically-thin gas and no turbulent or magnetic pressure support to the gas. However, we relax these assumptions one-by-one. In nature we may expect more than one of these effects to be present in the gas which will clearly introduce degeneracies but in the interests of presenting simple testable models we do not explore the entire parameter space. It is therefore important to consider this when devising observational strategies to test the model. We discuss the models and what observations may be robust tests of these models in the next section.

\section{Predictions and discussion}
\label{predictions}

Extended cool/cold gaseous filaments are observed in BCGs yet, in most cases, apparently have very low star formation efficiencies. 
The question of what may be preventing the gas from cooling - what is its excitation/heating mechanism - is an important one as it goes to the heart of how much can cooling hot X-ray gas affect late stage galaxy evolution. In this paper we hope to provide testable predictions for the `particle' excitation mechanism and to elucidate the importance of the surrounding hot gas in inhibiting the coldest/densest gas from forming and hence preventing star formation.

Making the assumption that the model put forward in F2009 and explored further here is the dominant excitation mechanism of the ionised and cold gas phases we can make predictions for further high spatial resolution cold gas observations of these filaments. The high spatial resolution of ALMA and HST is important as deriving gas masses and measuring the turbulent velocity of the filaments is highly dependent on the assumed gas fraction within the beam and on whether we are observing one or many filament threads.  In Section \ref{complicated} we look at the effects, on the \oisixthree and \ciionefiveseven line ratios in the model, of relaxing one-by-one some of the assumptions of F2009. Many mechanisms may be playing a part in setting the line ratios. However, applying Occam's razor we wish to single out a possible dominant mechanism.

Observations have found the ratios of \ciionefiveseven/H$\alpha$ is approximately 0.8, \oisixthree/\ciionefiveseven$<$1 and where it has been possible to measure the typically weak \oionefourfive\ emission line the ratio of \oisixthree/\oionefourfive$\sim$5-30 \citep{mittal2011, mittal2012, werner2013, werner2014}. As discussed in Sections \ref{additional_cold_clouds} the addition of a sink of very cold clouds at temperatures $\sim$10~K cannot explain the FIR line ratios while maintaining the ratios observed in the optical and NIR if the gas is all at constant pressure with the surrounding X-ray gas. \ciionefiveseven\ becomes optically thick at a greater column than \oisixthree\ (see Apendix \ref{optical_depth_appendix}) and the high \oisixthree/\oionefourfive\ line ratios also rule out these emission lines being optically thick and suggest they come from the warm 50$-$100~K gas (see Fig. \ref{oi_cii_depth}). Masing and foreground absorption in the [O {\sc i}] lines can change these ratios but would tend to lower rather than increase the ratio (see discussion in \citealt{liseau2006}). 

\begin{figure*}
  \begin{center}
   \includegraphics[width=0.49\textwidth]{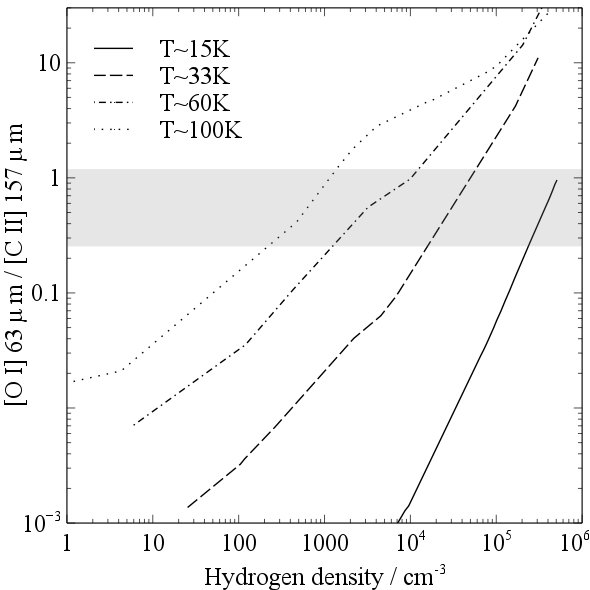} 
   \includegraphics[width=0.49\textwidth]{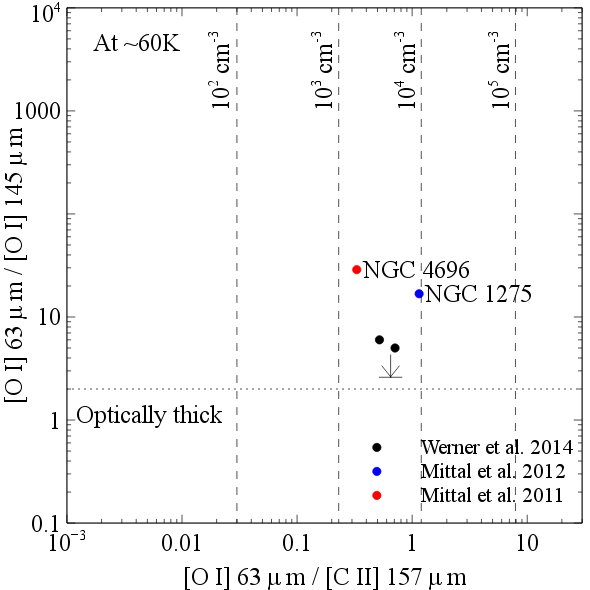} 
   \caption{{\bf Left:} If the gas is optically thin then at reasonable temperatures and densities the \oisixthree\ / \ciionefiveseven\ ratio traces fairly well the gas pressure. Therefore at constant temperature (or density) we can use the ratio as a density (or temperature) diagnostic. Here we plot the emission lines ratio against the density for clouds of different temperatures. The grey band indicates the detected line ratios in some BCGs and giant elliptical galaxies. {\bf Right:} A \oisixthree\ / \oionefourfive\ versus \oisixthree\ / \ciionefiveseven\ diagnostic diagram for a cloud at 60 K. \label{diagnostics}}
  \end{center}
\end{figure*}

Assuming the clouds maintain constant pressure with the hot ICM the most promising scenario would be to include an additional, density dependent, pressure term. Increasing the pressure support in this density dependent fashion, such as through an increase in the turbulence or density dependent magnetic fields pushes the ratio of both \oisixthree/H$\alpha$ and \ciionefiveseven/H$\alpha$ to lower ratios for a range of pressures and powerlaw indexes. An additional pressure support such as this would leave the pressure contribution from the hot gas essentially the same while requiring less thermal pressure from the dense gas, changing the line ratios that are predicted from cold, dense clouds while not altering the line ratios from the 10,000~K, `warm', gas. If turbulence contributes all the additional pressure support then 2-10~\kmps\ turbulence is required to reach the observed line ratios of approximately unity. This additional support decreases the density of the cold gas and could therefore also explain the longevity of the filaments.

Other indications exist that thermal pressure support alone is unlikely to be able to keep the filaments in BCGs stable for long periods of time. The periodicity observed in the star formation regions of the filaments of NGC 1275 would require turbulence of $\sim$17~\kmps\ if all the additional pressure support came from turbulence \citep{canning2014}. However, we note that the linear structure and long lifetimes suggest some pressure support likely comes from magnetic fields. Additionally, in radio-galaxies, quiescent, out-flowing cold clouds are often observed  (e.g. \citealt{nesvadba2010, combes2013, alatalo2014}), albeit at greater velocities than filaments from typical `mechanical AGN feedback' in BCGs. \cite{guillard2014} show that turbulent energy seeded by the AGN feedback could be responsible for the bright \ciionefiveseven\ emission lines observed in these sources.

Whilst the far infrared lines observed by Herschel offer important diagnostics of the gas properties, estimating the total mass in the very cold gas still requires an extrapolation of the model to the coldest emission lines $\lesssim20$~K. Most current observations of CO have observed the low J levels which are likely to be optically thick and therefore unable to constrain the gas masses. Future observations targeting optically thin CO lines will help answer the question of how much mass is concealed in the cold gas in massive galaxies and therefore the efficiencies of their star formation rates.

\subsection{Model predictions}

The model we are advocating in this paper is one where the excitation is dominated by collisions with energetic particles creating pockets of partially ionised regions interspersed with molecular regions in the filaments in a `swiss-cheese-like' fashion. The gas is supported also by magnetic fields or modest turbulence, which itself could be seeded by the impinging particles, which add pressure support to the gas, lowing the density and preventing immediate collapse into stars. In this section we outline tests of this model and diagnostics of the conditions of the gas.

\subsubsection{\oisixthree/\oionefourfive\ and \oisixthree/\ciionefiveseven}
Our models show, for optically thin emission, in low temperature gas $\lesssim$150 K the \oisixthree/\ciionefiveseven\ ratio traces relatively well the pressure. So for fixed temperature the ratio can be used as a density diagnostic. In the left hand panel of Fig. \ref{diagnostics} we provide a plot of the optically thin line ratios of \oisixthree/\ciionefiveseven\ against density for gas with temperatures ranging between 15~K and 100~K. The current values of the ratio from Herschel observations are indicated by the grey region. Taking the gas to be $\sim$60~K, consistent with the \oisixthree/\oionefourfive\ in the optically thin limit, in the right hand panel we plot the \oisixthree/\oionefourfive\ against the \oisixthree/\ciionefiveseven\ ratios and indicate the densities. Current observations, where at least two of these lines are detected, are overplotted. For gas under these conditions the current observations indicate the density should be a few 10$^{3}$~cm$^{-3}$. It is important to note these ratios are for the total galaxy fluxes which includes the denser and in some cases star forming interiors as well as the extended filaments.

\subsubsection{High J CO emission lines and high density gas tracers}
Observations of the high J (J$=$3-2 or higher) CO rotation ladder with ALMA will shed further light on the excitation mechanism and the importance of both turbulence and magnetic fields. Observing high J transitions are important for three key reasons. First, the high J transitions are insensitive to the low temperature integration limits (see Fig. \ref{PART_OI_CII}); second, they are less likely to be optically-thick than the low J CO lines; and third, the shape of the high J rotation ladder will enable a sensitive determination of the gas temperature; we might expect the gas temperature to differ to the dust temperature if the gas is excited by energetic particles as is seen in the molecular clouds in the nuclear disk of the Galaxy (e.g. \citealt{yusefzadeh2007}). 

The ratios of these high J CO lines to H$\alpha$ is also a sensitive test of the presence of gas with densities greater than 10$^{3}$ (see Fig. \ref{PART_states3}) which allows limits to be placed on the levels of additional pressure support allowed as the assumption of total pressure equilibrium therefore enforces constant density cooling at very low temperatures. High density lines like those of HCN and HCO$^{+}$ ($n\gtrsim10^{4}$~cm$^{-3}$) should not be observed in the extended non-star forming filaments if the additional pressure support is equivalent to 10~\kmps\ turbulence. \cite{salome2008} have detected HCN (3-2) in the central regions of NGC 1275 but so far no detection of very dense gas in the extended regions of filaments has been observed.
Fig. \ref{dense} shows the expected emissivities of CO J(3-2), CO J(5-4), HCN and HCO$^{+}$ for a range of ionising particle fluxes and gas densities, the emissivity in these lines is highly density dependent. Constraining the density of the gas allows us to identify the level of additional pressure support and whether the clouds remain in pressure equilibrium in the coldest $\lesssim20$~K gas. Additionally, high spatial resolution spectroscopy with ALMA will enable us to measure the turbulent support in the cold gas. Putting these together we can also elucidate the level of magnetic support required to the total pressure budget.

\subsubsection{Low J $^{12}$CO emission lines}
Optical depth estimates from the $^{12}$CO J(1-0) emission lines suggest this line is optically thick in the filaments of BCGs (\citealt{salome2003, salome2011}; note that $^{13}$CO and C$^{18}$O are likely to be optically thin in the filaments and as such may be good diagnostics of the model). Ratios of these lines with other detected emission lines are thus not easily compared to predictions from our model. However, these lines are likely thermalised and can be used as sensitive thermometers for the gas as the excitation temperature will be equal to the kinetic temperature. Combined with ratios of high J emission lines of CO which are sensitive tracers of the density (see Fig. \ref{dense}) constraints can be put on the physical conditions of the coldest gas.\\

We have argued above that collisional excitation with the hot particles in the surrounding X-ray gas may inhibit the dense conditions required for star formation in the extended filaments. However, we know of a few cases in which some filaments have disrupted into stars (e.g. \citealt{mcdonald2009, odea2010, canning2014}). In regions in which young stars are observed obviously emission lines dominated by excitation from young stellar sources will be observed. We might also expect the turbulence to be higher in these regions due to driving from young stellar winds hence the diagnostics suggested above are not relevant to regions which are actively forming stars and should not be applied to star forming knots in filaments. 

However, some information may still be gleamed from studying these regions. Additional pressure support of the gas, either by turbulence or by magnetic fields should alter the Jeans mass. Collisional excitation with energetic particles from the surrounding gas heats the gas and increases its ionisation fraction which will increase its coupling to magnetic fields, which should slow the timescale for gravitational collapse. We speculate that this excitation mechanism could therefore lead to a top heavy IMF in the filaments explaining the very massive star clusters observed in some outer filaments \citep{canning2014}.

\section{Conclusions}

We have extended the particle heating mechanism first presented in \cite{ferland2009} to examine the effects of abundant cold clouds, optically-thick gas, turbulence and magnetic fields on the bright line ratios from cold, $\lesssim$100 K, gas. We explore these models in the context of extended cool and cold gaseous nebulae observed in some massive galaxies many of which are devoid of star formation. We wish to be clear that we are not considering the centrally peaked emission and associated star forming regions observed in many massive galaxies where it is clear many process are occurring. We focus instead on `clean' non-star forming, extended nebulae. 

We suggest the simplest explanation for the discrepancy between the predicted [C {\sc ii}] and [O {\sc i}] line ratios may be that there is a small amount of additional pressure support in the cold gas from either turbulence (2-10 \kmps) and/or density dependent magnetic fields and we present predictions for line ratios and diagnostics of the gas temperatures and densities. We suggest that turbulence may be driven by the influx of energetic particles impinging on the cold gas, creating a higher level of ionisation in the gas and decreasing the gas density which could also inhibit star formation in these filaments. The majority of filaments would remain long lived unless, the gas pressure drops, star formation is either triggered by external perturbations, such as disturbances from rising AGN bubbles or galaxy interactions or the heat from such external perturbations is strong enough to eradicate the filaments.

\section{Acknowledgements}

REAC thanks E. Churazov for helpful and interesting discussions.
This work is based in part on observations made with Herschel,
a European Space Agency Cornerstone Mission with significant
participation by NASA. Support for this work was provided by
NASA through award number 1428053 issued by JPL/Caltech.
GJF acknowledges support by NSF (1108928, 1109061, and 1412155), NASA (10-ATP10-0053, 10-ADAP10-0073, NNX12AH73G, and ATP13-0153), and STScI (HST-AR- 13245, GO-12560, HST-GO-12309, GO-13310.002-A, and HST-AR-13914). The contour plots were produced with the {\sc veusz} plotting program (http://home.gna.org/veusz/). 

\bibliographystyle{mn2e}
\bibliography{mnras_template}
\appendix
\section{C{\sc loudy} code}

{\small

\noindent c

\noindent c

\noindent table HM05 z=0 

\noindent extinguish by 21, leakage = 0

\noindent cmb redshift 0

\noindent c

\noindent c

\noindent atom H2 levels large

\noindent atom H-like Lyman pumping off

\noindent abundances he =-1.022 li =-10.268 be =-20.000 b =-10.051 c =-3.523 n =-4.155

\noindent continue o =-3.398 f =-20.000 ne =-4.222 na =-6.523 mg =-5.523 al =-6.699

\noindent continue si =-5.398 p =-6.796 s =-5.000 cl =-7.000 ar =-5.523 k =-7.959

\noindent continue ca =-7.699 sc =-20.000 ti =-9.237 v = -10.000 cr = -8.000 mn =-7.638

\noindent continue fe = -5.523 co =-20.000 ni =-7.000 cu =-8.824 zn =-7.6990 no grains

\noindent grains ism

\noindent grains pah

\noindent set pah constant -4.6

\noindent set H2 Jura rate

\noindent case B

\noindent c

\noindent c

\noindent cosmic rays background -1 vary

\noindent grid from 0 to 7 in 0.1 dex steps

\noindent hden 4 vary

\noindent grid 0 6 0.1

\noindent c

\noindent c

\noindent stop zone 1

\noindent set dr 0

\noindent stop temperature off

\noindent turbulence = 2 km/s}

\section{Optical depths}
\label{optical_depth_appendix}

\begin{figure*}
  \begin{center}
   \includegraphics[width=0.4\textwidth]{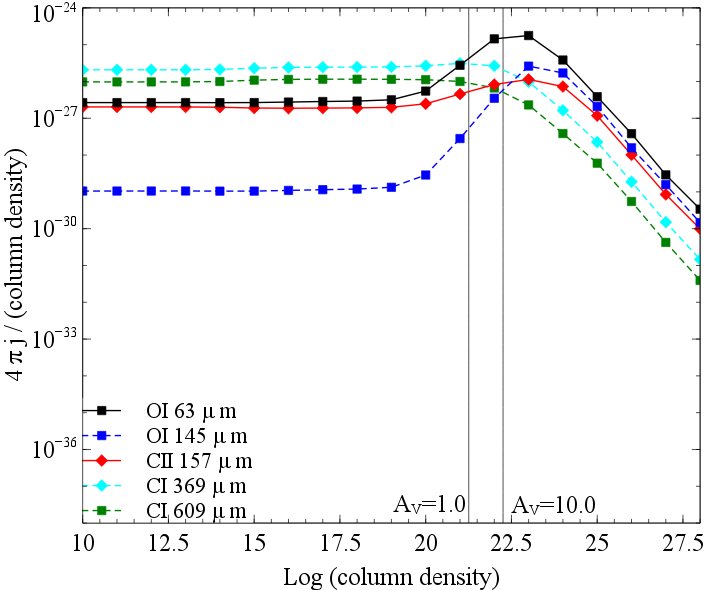}    \includegraphics[width=0.4\textwidth]{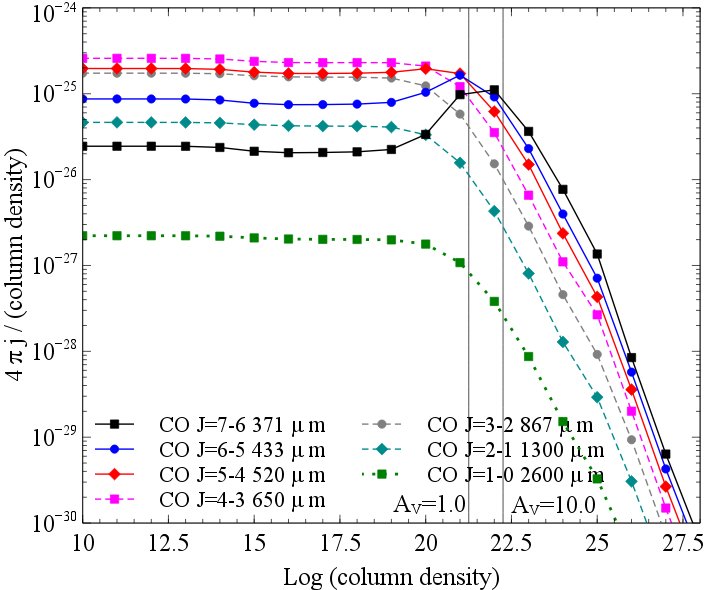}    \includegraphics[width=0.4\textwidth]{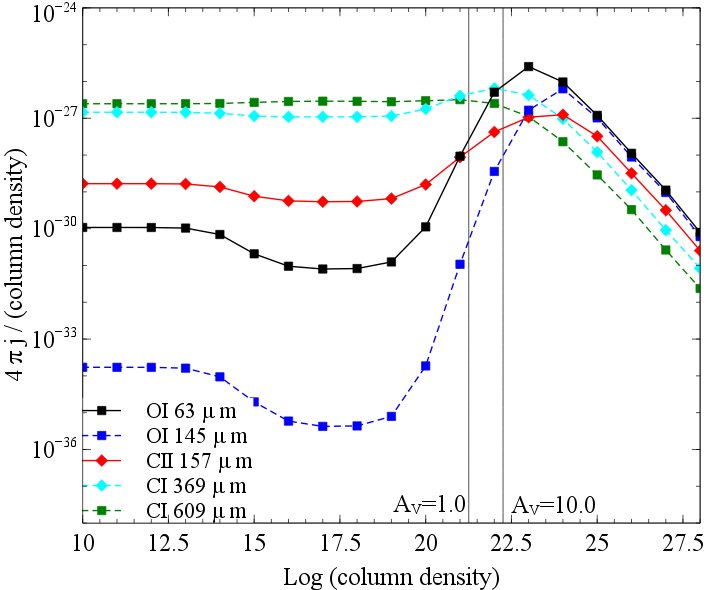}    \includegraphics[width=0.4\textwidth]{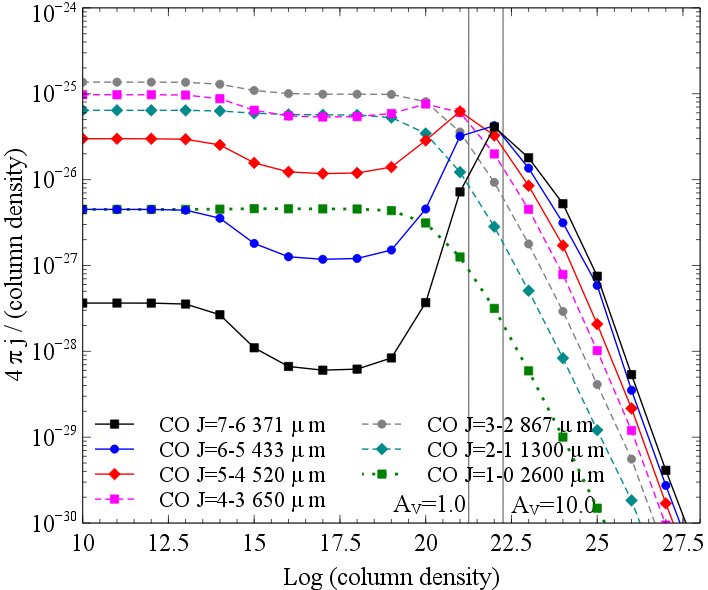}    \caption{The predicted line emissivity divided by the column density plotted against the column density of the clouds for clouds at constant pressure. The surface temperature of the cloud in the upper plot is T$\sim$33~K while the lower plot has a lower surface temperature of T$\sim$15~K. \label{emissivity_column_const_press}}
  \end{center}
\end{figure*}

In Fig. \ref{emissivity_column_const_press} we shown the emission line emissivity divided by the cloud column density against the cloud column density for two models.  The first has a hydrogen density of 10$^{5.3}$ cm$^{-3}$ and a particle density of 10$^{2}$ times that of the Galactic background cosmic ray energy density (the Galactic background is here taken as 1.8 eV cm$^{-3}$) which results in a cloud surface temperature of $\sim$15~K. We choose this temperature and density model as an optically-thin gas under these conditions has \oisixthree\ over \ciionefiveseven\ less than one. The second has a hydrogen density of 10$^{5.0}$ cm$^{-3}$ and a particle density of 10$^{2.5}$ times that of the Galactic background cosmic ray energy density leading to a higher temperature, $\sim$33 K. This is close to the temperature at which the emission expected from both lines is equal for an optically-thin gas (see right hand panel of Fig. \ref{PART_OI_CII}). Both models have a pressure of $\sim10^{6.5}$~\pcmcuK. 

For an optically-thin line the emissivity will increases in proportion to the column, this produces horizontal lines on the plots. The column density at which the column begins to increase faster than the emissivity indicates the column at which the line has become optically-thick and line photons are collisionally dexcited following multiple scatterings. The [O I]63$\mu$m line in the shown model (upper left plot), becomes optically-thick around a column density of $\sim10^{22}$~cm$^{-2}$, corresponding to a high extinction of A$_{\mathrm{V}}=$5.6, assuming N$_{\mathrm{H}}$=1.8$\times$10$^{21}$A$_{\mathrm{V}}$ \citep{predehl1995}. The \oionefourfive and \ciionefiveseven emission lines become optically thick at columns greater than $\sim10^{22.5}$~cm$^{-2}$ which are very rare and so these lines are likely optically thin.

In the right hand panels we show the emissivities in the CO rotation ladder as a function of column density. The low J lines of CO have large optical depths in gas with low extinction A$_{\mathrm{V}}<$1 and are most likely optically thick. These lines may not be constraining to test the various excitation models of the filaments but can be sensitive thermometers of the gas temperature.

\section{Emissivities}

\begin{figure*}
  \begin{center}
    \includegraphics[width=0.33\textwidth]{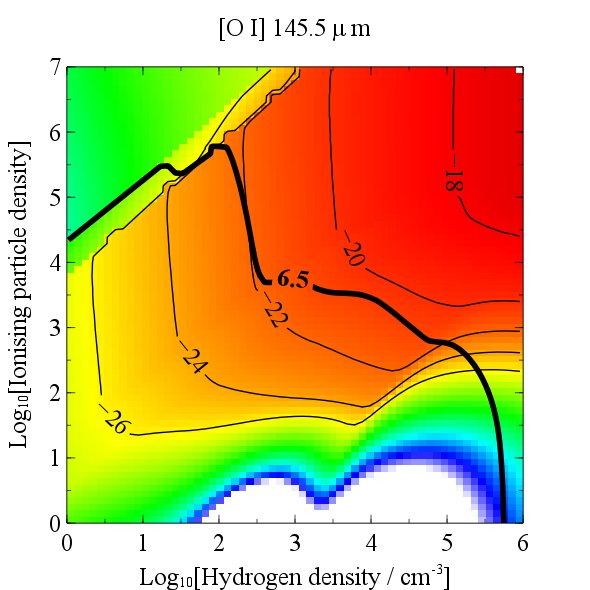}     
   \includegraphics[width=0.33\textwidth]{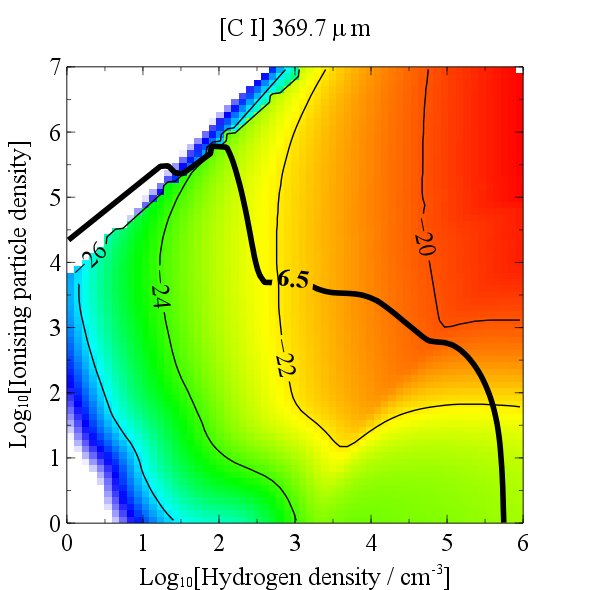}    
   \includegraphics[width=0.33\textwidth]{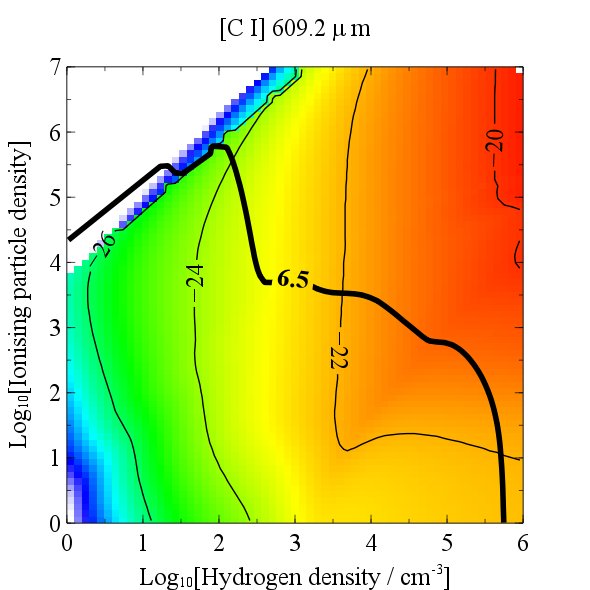}    
   \includegraphics[width=0.4\textwidth]{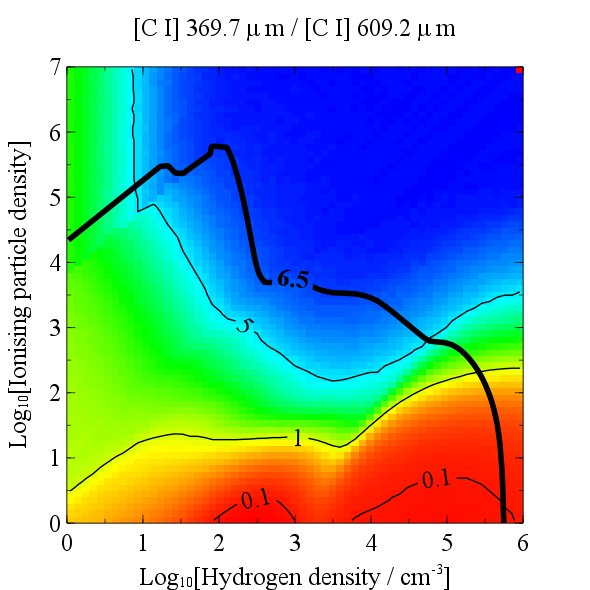}     
     \includegraphics[width=0.4\textwidth]{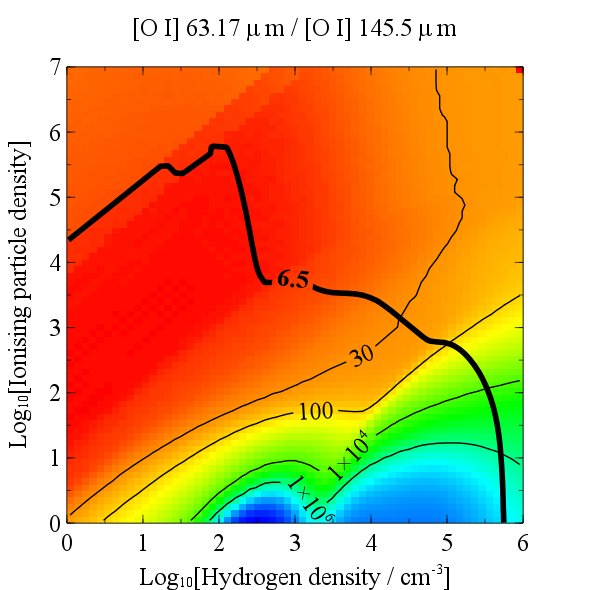}         
\caption{Predicted log emmisivities of the \oionefourfive, \cithreesixnine, and \cisixzeronine\ transitions and the neutral carbon and neutral oxygen line ratios for an optically thin gas. The countour indicates a gas pressure of 10$^{6.5}$~\pcmcuK. \label{other_em}}
  \end{center}
\end{figure*}

\begin{figure*}
  \begin{center}
   \includegraphics[width=0.4\textwidth]{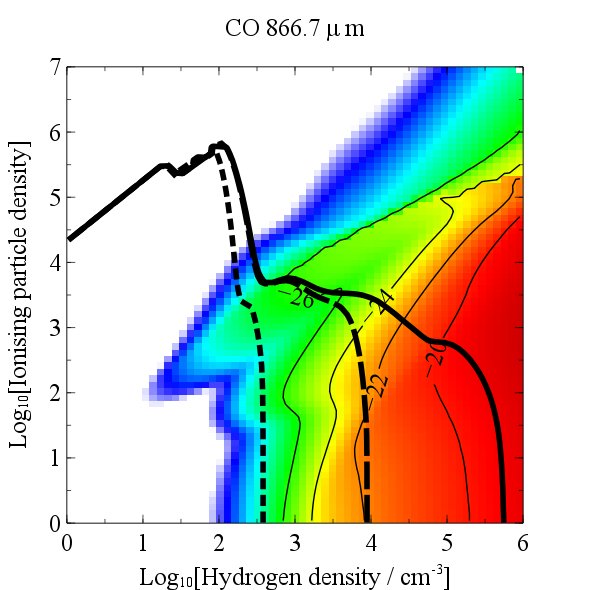}    
   \includegraphics[width=0.4\textwidth]{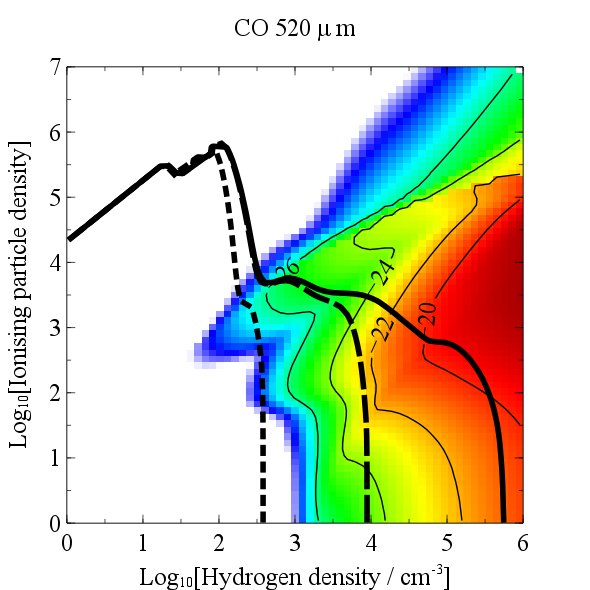}    
   \includegraphics[width=0.4\textwidth]{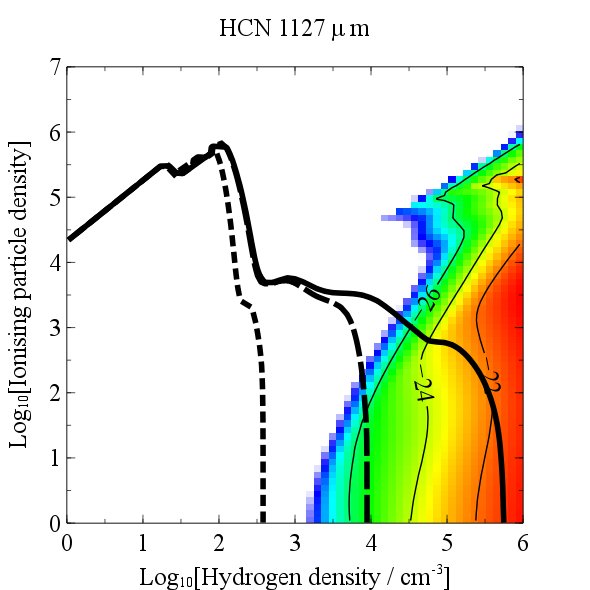}     
   \includegraphics[width=0.4\textwidth]{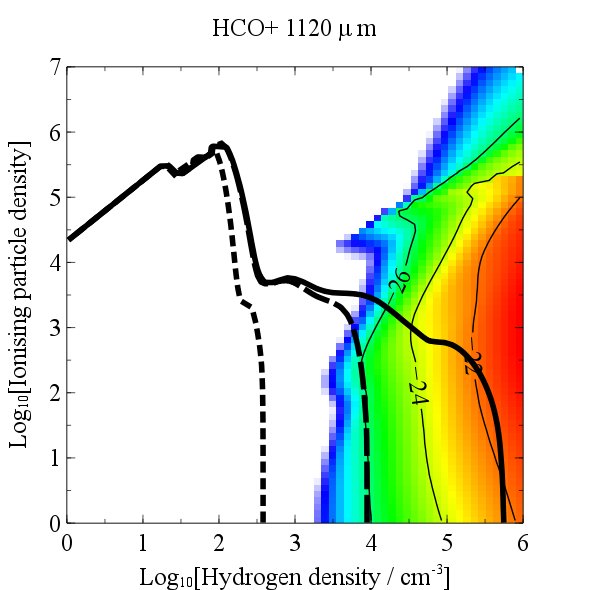}    
\caption{The predicted log emissivities of the CO J(3-2) and J(5-4) transitions and those of lines of HCN and HCO$+$ overlaid with contours of the gas pressure as shown in Fig. \ref{turb_figs}. All these lines offer sensitive diagnostics of the gas density and the CO ladder transitions can also offer sensitive temperature measurements. If a high degree of additional pressure support is present in the filaments than the densest gas should not be. Observations of the strength of these density sensitive lines in regions in which no star formation is occurring can constrain the additional pressure support in the gas.  \label{dense}}
  \end{center}
\end{figure*}

For completeness in Fig. \ref{other_em} and \ref{dense} we show the optically thin emissivities of other key strong lines from our model and the ratios for the neutral carbon and oxygen lines. In the cold gas the optically thin [O {\sc i}] line ratios are a relatively faithful indicator of the gas temperature while the [C {\sc i}] ratios are more sensitive to the particle ionising flux. The lines shown in Fig. \ref{dense} are good indicators of the gas density. 


\end{document}